# IDP-PGFE: An Interpretable Disruption Predictor based on Physics-Guided Feature Extraction

Chengshuo Shen, Wei Zheng*, Yonghua Ding, Xinkun Ai, Fengming Xue, Yu Zhong, Nengchao Wang, Li Gao, Zhipeng Chen, Zhoujun Yang, Zhongyong Chen, Yuan Pan and J-TEXT team†
International Joint Research Laboratory of Magnetic Confinement Fusion and Plasma Physics, State Key Laboratory of Advanced Electromagnetic Engineering and Technology, School of Electrical and Electronic Engineering, Huazhong University of Science and Technology, Wuhan, 430074, China
**Email**:
shenchengshuo@hust.edu.cn, zhengwei@hust.edu.cn

## Abstract

Disruption prediction has made rapid progress in recent years, especially in machine learning (ML)-based methods. If a disruption prediction model can be interpreted, it can tell why certain samples are classified as disruption precursors. This allows us to tell the types of incoming disruption for disruption avoidance and gives us insight into the mechanism of disruption. This paper presents a disruption predictor called Interpretable Disruption Predictor based on Physics-Guided Feature Extraction (IDP-PGFE) and its results on J-TEXT experiment data. The prediction performance of IDP-PGFE with physics-guided features is effectively improved (TPR = 97.27%, FPR = 5.45%, AUC = 0.98) compared to the models with raw signal input. The validity of the interpretation results is ensured by the high performance of the model. The interpretability study using an attribution technique provides an understanding of J-TEXT disruption and conforms to our prior comprehension of disruption. Furthermore, IDP-PGFE gives a possible mean on inferring the underlying cause of the disruption and how interventions affect the disruption process in J-TEXT. The interpretation results and the experimental phenomenon have a high degree of conformity. The interpretation results also gives a possible experimental analysis direction that the RMPs delays the density limit disruption by affecting both the MHD instabilities and the radiation profile. PGFE could also reduce the data requirement of IDP-PGFE to 10% of the training data required to train a model on raw signals. This made it possible to be transferred to the next-generation tokamaks, which cannot provide large amounts of data. Therefore, IDP-PGFE is an effective approach to exploring disruption mechanisms and transferring disruption prediction models to future tokamaks.

† See the author list of "N. Wang *et al* 2022 Advances in physics and applications of 3D magnetic perturbations on the J-TEXT tokamak, *Nucl. Fusion* **62** 042016"

## 1. Introduction

The safe and stable operation of next-generation tokamaks requires effectively avoiding and mitigating disruptions [1,2,3], which relies on predictions and in-depth understanding of plasma instabilities and disruption. The complexity of the physical mechanism of disruption leads to the difficulty in using first-principle models to predict plasma instabilities and disruptions [4]. Physics-driven and data-driven are two main approaches to study disruption prediction. On the one hand, physics-driven disruption prediction studies [5–7] put effort into identifying disruption or instabilities event chains for operators to avoid disruptions. Such as, the disruption event characterization and forecasting (DECAF) [8] suite contains first-principle, physics-based modules for instabilities identification in tokamaks. On the other hand, most recently developed disruption predictors are based on data-driven methods because of the better performance [9–13].

However, most of these data-driven models do not necessarily reflect the dynamics behind the phenomenon. Their physics fidelity and the interpretability of their results require in-depth investigation. Interpretable disruption prediction research has made preliminary progress these years. The physics-driven models try to combine the advantages of both paradigms by adopting surrogate machine learning (ML) models [14], which could also improve the interpretability of these ML models. In data-driven method studies, an interpretability model, achieved by applying symbolic regression methods [15–17], has been obtained with the support vector machine (SVM) in JET. An approach to interpret the 1.5D convolutional neural network (CNN) model has been developed in HL-2A [18], which is a counterfactual-based interpretable approach. A Class Activation Mapping method has also been used to interpret the deep learning model predictions [19]. The hazard-based event prediction method has been introduced on DIII-D [20]. Its full deployment requires additional assumptions (models) about the future evolution of the relevant covariates. The physics-based indicators of disruption precursors based on the peaking factors were developed and adopted at JET[21–23]. They have been also implemented to interpret machine learning-based disruption predictors across tokamaks on DIII-D and JET [24]. Moreover, and automatic feature extraction based on deep learning from plasma profiles was also developed [25].

The interpretable disruption prediction research is in the infancy stage. In comparison, the interpretability research has made more advanced progress in machine learning research. A prototypical part network (ProtoPNet) [26] dissects the image by finding prototypical parts to construct a deep interpretable network architecture. The perspective that guided the interpretability model design is that we should stop "interpreting" decisions made by black-box models after the fact and instead build models that are constructively interpretable [27]. This kind of perspective could also be suitable for disruption prediction and inspire us that we should design a disruption predictor that is inherently interpretable. In mathematics, machine learning-guided intuition helps researchers to prove theorem via attribution techniques [28]. SHapley Additive exPlanations (SHAP) is a unified attribution-based interpretable approach [29–31], which has already been successfully applied in explaining the prediction of hypoxemia during surgery [32].

The perspective and techniques in machine learning research could give us more inspiration for interpretable disruption prediction research. The focus of most disruption predictors is accuracy or cross-machine capability. In this work, we would like to design a disruption predictor on the purpose for both accuracy and interpretability. In our previous work, SHAP was applied to the LightGBM [33] based disruption predictors for J-TEXT and HL-2A [34]. The interpretability study provided a strategy for selecting diagnostics and shots data for developing cross-machine predictors. However, the interpretability of the model is not yet in-depth enough due to the limitations of the input signals, the raw diagnostic signals. The work from DIII-D and JET teams [21,23,24] found that physics-based disruption markers in data-driven algorithms are a promising path toward realizing a uniform framework to predict and interpret disruptive scenarios across different tokamaks. An interpretable disruption predictor could also be designed based on physics-based disruption markers. Therefore, in this work, we would like to design an interpretable disruption predictor based on physics-based disruption markers by using SHAP as the interpretable approach.

The way to obtain the physics-based disruption markers in our work is called Physics-Guided Feature Extraction (PGFE). The aim of PGFE is not only to enable the model interpretability but also to improve the predictor's accuracy. Deep learning's powerful feature extraction capabilities make it typically outperform traditional machine learning when using a large enough amount of raw signals. However, when there is not enough data, inductive bias can be introduced to the network to better embedded known knowledge of disruption in the model. Inductive bias will limit the search domain of the machine learning algorithm, so it can get a better result when there is not enough data. Using feature engineering with known knowledge of disruption to reduce the feature space dimension limits the search domain of the training algorithm so that it can be considered as adding inductive bias. Adding domain knowledge to machine learning is thought necessary to reach good performances in low data regimes [35], and it is the base for meaningful physical interpretation. PGFE could even reduce the data requirement of the predictor. Due to the high cost of disruptive data from ITER and next-generation tokamaks, disruption prediction researchers need to find a way to cross-machine prediction or reduce the requirement of disruptive data. Cross-machine has been studied through deep learning [36–38] and adaptive learning [39]. Inductive bias in PFGE is specific to disruption prediction tasks. Therefore, training the model with less data is accessible. It is possible to achieve a disruption prediction model of the new device directly using a small amount of discharge data with PGFE.

In this paper, an Interpretable Disruption Predictor based On PGFE (IDP-PGFE) has been introduced. The purpose of this work and the advantages of PGFE have already been described in this section. The following section will describe the structure of IDP-PGFE, which consists of a feature extractor, a disruption classifier, and an explainer. Section 3 introduced the dataset on J-TEXT. The predictive performance of IDP-PGFE on J-TEXT is followed in section 4, which shows the two advantages of PGFE. Section 5 examines the interpretability of the predictor, including its application to the J-TEXT physics experiment. Section 6 will briefly discuss the potential and limitations of IDP-PGFE. The summary is in section 7.

## 2. The three components of IDP-PGFE: feature extractor, disruption classifier and explainer

IDP-PGFE consists of three components, feature extractor, disruption classifier and explainer. This section will describe these three components of the IDP-PGFE.

### *2.1 Feature extractor: PGFE*

The diagnostics signals in the tokamak experiment are heterogeneous, meaning not all dimensions share the same physical meaning, and the relation among them is very complex. It is tough to get physically meaningful patterns from the interpretation results. PGFE is a "white-box" which extracts features with physical meanings for in-depth study after the explainer.

Multiple physical phenomena that may be part of a chain of disruption [5] can be observed in tokamaks. In this work, the physics-guided features have been extracted based on MHD instabilities, radiation, density related disruption and basic plasma control system (PCS) signals. The descriptions of features and diagnostics for extracting them can be found in table 1. The numbers followed the name of diagnostic in the third column are the number of channels used for feature extraction. MHD instability is a significant precursor to disruption, especially the locked mode [5]. Moreover, 2/1 magnetic island growth [40] and multi-magnetic island overlap [41,42] are also the possible precursors of disruption. This work uses Mirnov probes and locked mode detectors to extract this type of feature, as listed in the first row of Table 1.

Table 1 Descriptions and symbols of all the features

| Types of features | Relation to disruption | Channels of diagnostics | Symbol |
|---|---|---|---|
| MHD instabilities related | 2/1 magnetic island growth; Multi-magnetic island overlaps; Locked mode | Mirnov probe in poloidal array | Mir_abs<br>Mir_fre<br>Mir_Vpp |
| | | Mirnov probes in poloidal array (2) | mode_number_m |
| | | Mirnov probes in toroidal array (2) | mode_number_n |
| | | Locked mode detectors (4) | n=1 amplitude |
| Radiation related | Temperature hollowing; Edge cooling; | soft x-ray (SXR) array (30) | SXR_array_kurt (skew, var) |
| | | Central channel of SXR array | SXR_core |
| | | Central channel of CIII radiation array | CIII |
| | | Absolute eXtended Ultra Violet (AXUV) array (30) | AXUV_array_kurt (skew, var)<br>P_rad |
| Density related | Density limit | Far-infrared three-wave polarimeter-interferometer (FIR) array (17) | FIR_array_kurt (skew, var)<br>sum_$n_e$ |
| | | Central channel of FIR array | $n_{e0}$ |
| Basic PCS signals | Plasma out of control; | Toroidal field and plasma current | $B_t$, $I_p$ |
| | | Horizontal and vertical field current ($I_{hf}$ and $I_{vf}$), | $I_{hf}$, $I_{vf}$, |
| | | Difference of $I_{hf}$ and $I_{vf}$ | $I_{hf}$_diff, $I_{vf}$_diff |
| | | Horizontal and vertical displacements, | $d_r$, $d_z$ |

Temperature hollowing and edge cooling [43] are risk indicators for MHD instabilities, and they could also be treated as the disruption precursors. Two radiation arrays, soft *x*-ray (SXR) [44] and Absolute eXtended Ultra Violet (AXUV) [45] arrays, are used to extract this type of features due to the difficulty in accessing temperature profiles in all the discharges in the dataset. Theoretically, density limit disruption could be predicted through Greenwald density limit [46]. However, density limit disruption usually does not reach the Greenwald density limit in J-TEXT. Therefore, the scaling law is not suitable for predicting disruptions in J-TEXT density limit disruption. Density related precursors could be extracted through far-infrared three-wave polarimeter-interferometer (FIR) [47]. The others are basic PCS signals, where the differential of the horizontal and vertical currents can effectively reflect the plasma control information and is extracted as features. As a result, 92 channels of diagnostics have been extracted into 28 features. Details of MHD instabilities, radiation and density related features extraction engineering is described.

### *2.1.1 MHD instabilities related features*

In J-TEXT, the growth of $m/n$ = 2/1 tearing mode (TM) and multi-magnetic islands overlap are the possible precursors of disruption. Mirnov probes and locked mode detectors are the fundamental diagnostics for MHD instabilities measurement in J-TEXT [48], which provide a simple, robust measurement of static and fluctuating magnetic properties. Deep learning approach to process the Mirnov coil signal [19] and manual approach to the locked mode signal [49] have also been developed at JET.

"*Mir_Vpp*" is the peak-to-peak value of the Mirnov probe after a low-pass filter with a cut-off frequency of 50 kHz. The typical frequency of 2/1 tearing mode in J-TEXT is approximately 3 kHz. A sliding window of 5 ms (longer than 5 periods of TM rotation) is selected as the time window of the Fast Fourier Transform (FFT) on the Mirnov probe. $X^i(f)$ represents the Fourier transform of the $i^{th}$ time window (slice). "*Mir_abs*" and "*Mir_fre*" are the intensity and frequency of $X^i(f)$. If multiple frequencies exist in the Mirnov probe frequency spectrum, the one with the highest spectral intensity will be selected. Here, we did not use the integral Mirnov signals to avoid the zero-drift uncertainty between different discharges that contributed to our prediction. $X_1{}^i(f)$ and $X_2{}^i(f)$ represent the Fourier transform of two Mirnov probe signals of each slice. Their cross spectral density (CSD) can be expressed as

$$P_{12}^{j}(f)=X_{1}^{j*}(f)X_{2}^{j*}(f)=A(f)e^{i\delta_{12}^{j}(f)} \tag{2-1}$$

where $P_{12}{}^i(f)$ is the cross spectral density between two Mirnov probes, $A(f)$ is the absolute value of $P_{12}{}^i(f)$, and $\delta_{12}{}^i(f)$ is the phase of $P_{12}{}^i(f)$. If the two Mirnov probes are in the poloidal array, the mode number $m$ can be calculated as

$$m=\frac{\delta_{12}^{j}(f)}{\theta} \tag{2-2}$$

where $\theta$ is the poloidal separation between the two Mirnov probes. In J-TEXT, $m$ = 3 and $m$ = 2 modes often lead to a multi-magnetic island caused disruption. Hence, the "*mode_number_m*" (*MNM*) uses a weighted average mode number indication, which is described as (2-3)

$$<m>=\frac{\sum\delta_{12}^{j}(f_k)A(f_k)}{\theta\sum A(f_k)} \tag{2-3}$$

where $k$ represents the different frequencies of the CSD. The "*mode_number_n*" (*MNN*) is also calculated like this. To increase the accuracy of *MNM* and *MNN*, only the frequency component with a coherence larger than 0.95 will be considered.

A typical discharge with TM of J-TEXT is shown in Figure 1. The evolution of *MNM*, *MNN*, *Mir_fre*, and *Mir_abs* with the raw Mirnov signals are shown. Because of the weighted average mode number indication, *MNM* and *MNN* are continuous. The slowdown of mode frequency and growth of mode amplitude induced disruption after the short duration locked mode (less than 1ms). The $m/n$ of this mode is 2/1 and the *Mir_fre* is approximately 3 kHz before mode locking. The ramp down of *Mir_fre* represents the mode locking.

After mode locking, *n = 1 amplitude* is also the MHD-related feature. The magnetic field measured by the locked mode (LM) detector can be expressed as

$$\mathrm{B}_r(\theta,\varphi)=\sum b_r^{n=i}\cos(m\theta+n\varphi+\xi^{n=i}) \tag{2-4}$$

where $\theta$ and $\varphi$ are the poloidal and toroidal location of the LM detector, $m$ and $n$ are the poloidal and toroidal mode number, and $\xi$ is the spiral phase. If only consider the main component in J-TEXT, which are $n$ = 0, 1 and 2, for the detector on the middle plane $\theta$ = 0 (The following are all based on this situation), equation (2-4) can be expressed as

$$\mathrm{B}_r(\theta=0,\varphi)=b_r^{n=0}+b_r^{n=1}\cos(\varphi+\xi^{n=1})+b_r^{n=2}\cos(2\varphi+\xi^{n=2}) \tag{2-5}$$

*n = 1 amplitude*, $b_r{}^{n=1}$ can be calculated through two LM detectors with $\Delta\varphi = \pi$, which is shown in equation (2-6)

$$b_r^{n=1}\cos(\varphi+\xi^{n=1})=\frac{B_r(\varphi)-B_r(\varphi+\pi)}{2}\,. \tag{2-6}$$

Then $b_r{}^{n=1}$ and $\xi^{n=1}$ can be calculated by fitting two pairs of these LM detectors. The application of *n = 1 amplitude* in the resonant magnetic perturbation (RMP) experiments has been introduced in our review paper on magnetic diagnostics [48].

In the limiter configuration with a circular cross-section of J-TEXT plasma, the features "*Mir_abs*", "*Mir_fre*" and "*Mir_Vpp*" are extracted through one Mirnov probe; "*mode_number_m*" (*MNM*) and "*mode_number_n*" (*MNN*) are extracted from two Mirnov probes in poloidal and toroidal array, respectively; and "*n = 1 amplitude*" is extracted from two pairs of LM detectors, each pair consist two locked mode detectors with the toroidal angle of 180°.

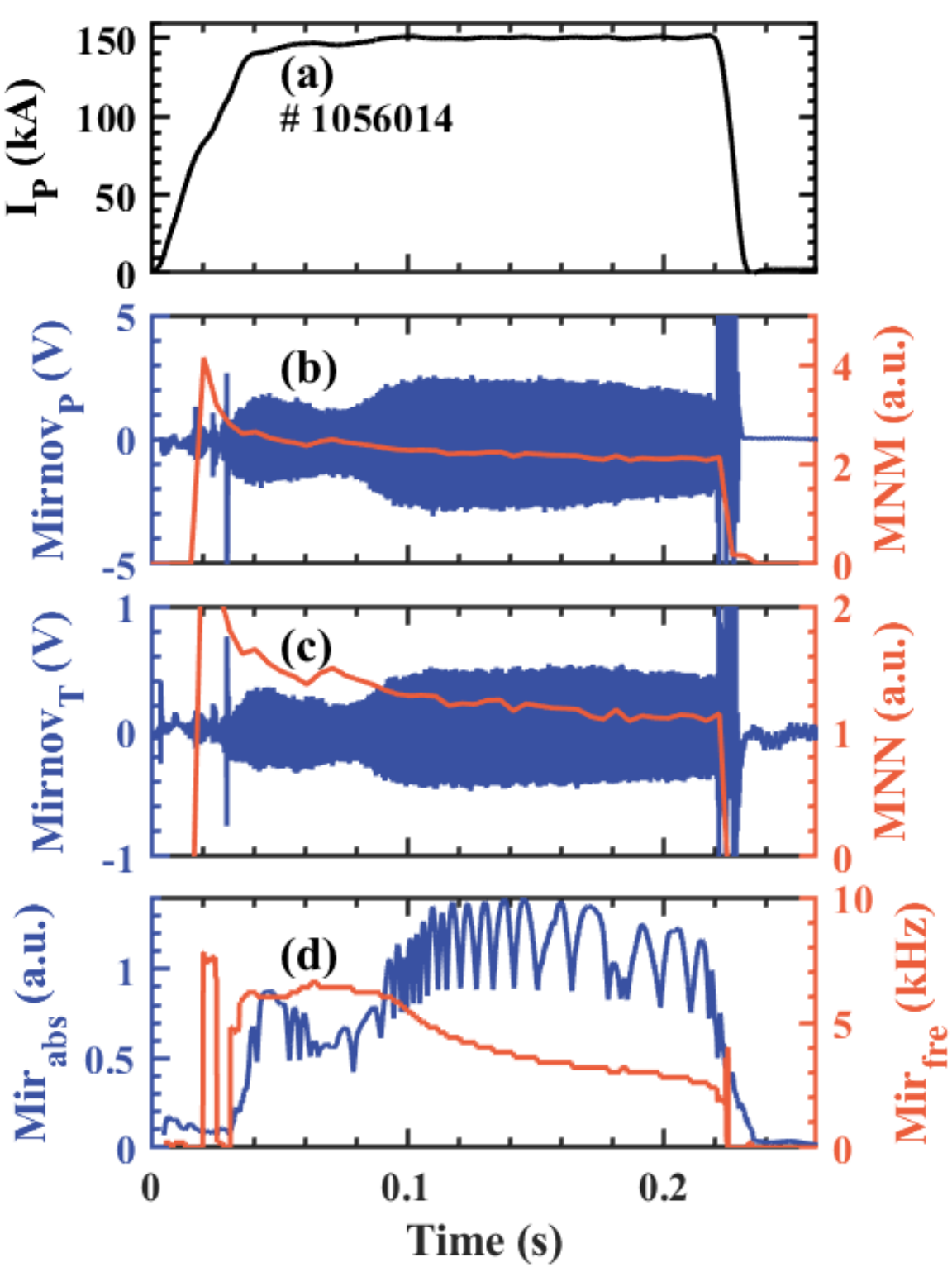

Figure 1 A typical disruptive discharge with tearing mode of J-TEXT. a) shows the plasma current of the discharge; b) and c) are the raw signal (left) and mode number (right) of the poloidal and toroidal Mirnov signals; d) shows *Mir_abs* (left) and *Mir_fre* (right) of one Mirnov signal (the raw signal in b)).

### *2.1.2 Radiation and density related features*

Temperature hollowing and edge cooling are phenomena of performance in profiles, which are 1D signals. In order to include the radial profiles of 1D signals in data-driven analyses, 0D peaking factor metrics have been synthesized on JET and DIII-D [21,24]. Profile indicators [22,23,39] have also been applied on the transfer of adaptive predictors in JET and ASDEX Upgrade (AUG). Both the two methods are calculated the first order statistics, i.e. mean of the arrays. Deep learning models on JET and DIII-D can extract 1D signals to features through convolution [25,36], which includes higher dimensional information. In this work, we calculated the higher-order statistics (HOS) of the 1D signals from the SXR, AXUV and FIR arrays for radiation and density related features.

Two HOSs have been selected to extract the 1D signals to 0D features: skewness (*skew*), and kurtosis (*kurt*). Variance (*var*) is a low-order statistic. As the variance (second order) is higher order than arithmetic mean (first order), in this work the HOS is also included variance. Variance is a measure of dispersion, meaning it is a measure of how far a line of sight of the array is spread out from their average value. The feature, *var* is defined by:

$$var = \frac{1}{n}\sum_{k=1}^{n}(x_k - \mu)^2 \qquad (2\text{-}7)$$

where $n$ is the number of lines in the array, $x_k$ is the signal at the $k^{th}$ line of sight in the array, $\mu$ is the mean of all signals in the array. The plasma profiles are parabolic in the distribution in the limiter configuration with a circular cross-section of J-TEXT plasma. Higher *var* represents the greater span of data in the profile indicated a more peaked profile.

Skewness is a measure of the asymmetry of the distribution of the array signals about their mean. The feature, *skew* is defined by:

$$skew = \frac{\frac{1}{n}\sum_{k=1}^{n}(x_k - \mu)^3}{[\frac{1}{n}\sum_{k=1}^{n}(x_k - \mu)^2]^{3/2}} \qquad (2\text{-}8)$$

If $skew > 0$, the kind of skewness is called positive skew or right skewed state, which represents the mean > median > mode (statistics meanings) of the array signals. If $skew < 0$, then the opposite and if $skew = 0$, the three values above are the same.

Kurtosis is a measure of the combined weight of a distribution's tails relative to the centre of the distribution. The feature, *kurt* is defined by:

$$kurt = \frac{\frac{1}{n}\sum_{k=1}^{n}(x_k - \mu)^4}{[\frac{1}{n}\sum_{k=1}^{n}(x_k - \mu)^2]^2} - 3 \qquad (2\text{-}9)$$

"Minus 3" gives a kurtosis of 0 to the normal distribution. This kind of kurtosis is also called excess kurtosis, simplified to kurtosis in this paper. Higher *kurt* indicates fewer larger or smaller outliers and more concentrated data distribution in the profile. Therefore, higher *kurt* represents a fatter profile, while lower *kurt* represents a thinner profile.

The HOS features in one discharge are shown in figure 2. The blue line represents the 0D signal at the middle channel of each array, the green, yellow, and orange lines represent *kurt*, *skew*, and *var*, respectively. In this discharge, the increase of plasma radiation leads to disruption. The increase in *var* of SXR array and AXUV array represents the radiation profile gradually peaking. The slight reduction on the *kurt* of the FIR array indicates the gradient of density profile increased. The anomalous fluctuations in the *var*, *skew* and *kurt* before 0.15ms result from the plasma's horizontal displacement of the plasma not yet fully stabilised.

The detailed HOS feature of this discharge are shown in figure 3, ranging from 0.315s to 0.355s. Before 0.341s, the HOS features changed little. The *var* of each array increased, and the *skew* and *kurt* of each array decreased a little. After 0.341s, the HOS features show great variations and oscillations with the disappearance of the sawtooth. The *var* of FIR and SXR array and *skew* of FIR array decreased significantly. While the *kurt* of FIR array increased significantly and the var of AXUV array increased then decreased. Other HOS features show oscillations with slight variations. At about 0.349s, the disruption occurred.

Radiation related features also included the radiated power *P_rad* and 0D signals, such as impurities radiation *CIII* (carbon wall for J-TEXT) and soft x-ray radiation *SXR_core* at the middle channel. Density related features also included $sum_n_e$ and 0D signal core line integral density $n_{e0}$, line average density of the middle channel. The feature $sum_n_e$ is calculated by summing up all the signals in FIR array to express the electron density of the poloidal profile approximately.

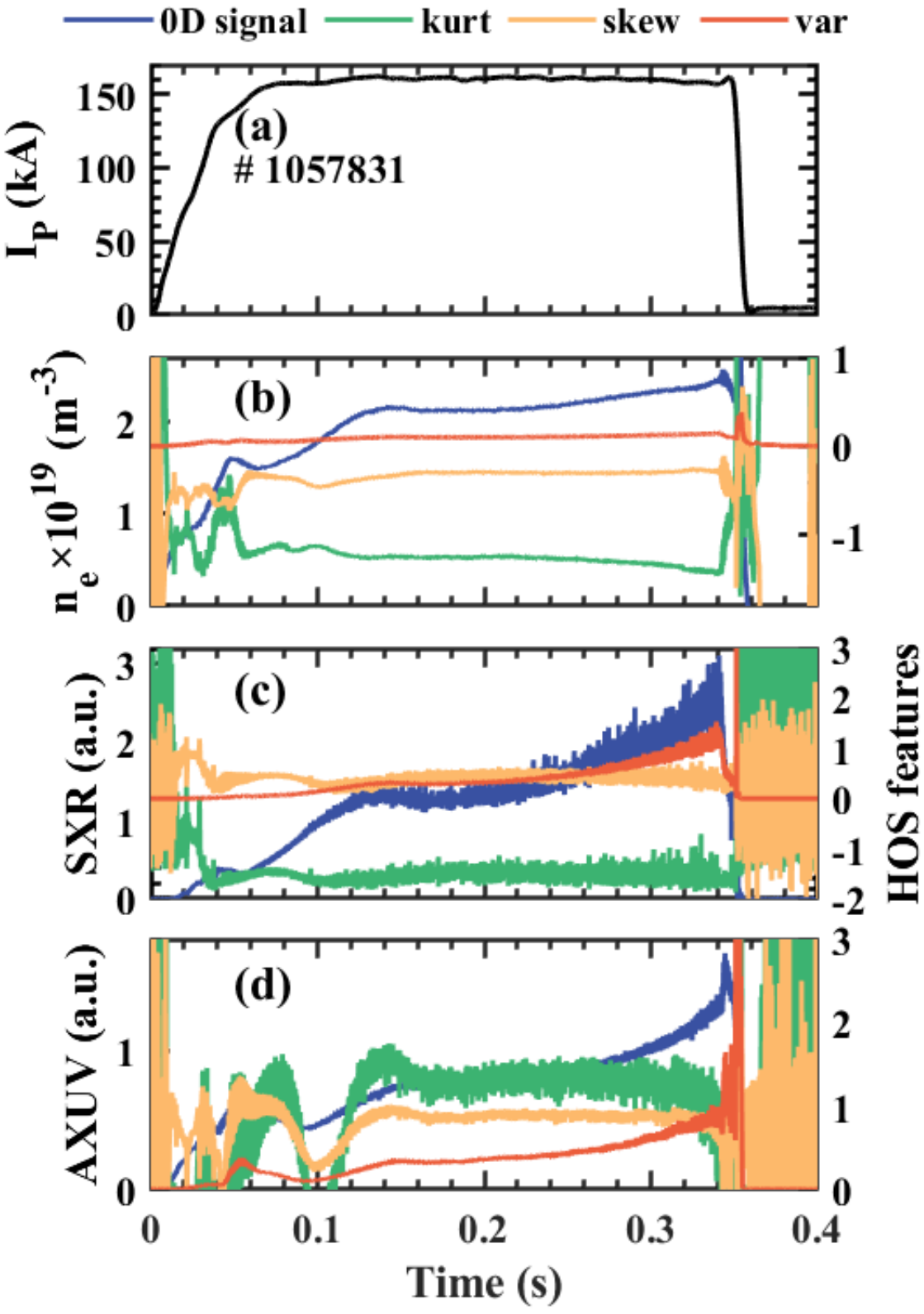

Figure 2 A disruptive discharge to show the HOS features in J-TEXT. The navy-blue line represents the 0D signal at the middle channel of each array, the green, yellow, and orange lines represent *kurt*, *skew*, and *var*, respectively. a) shows the plasma current of the discharge; b) shows the FIR array, which provided density signals; c) shows the SXR array and d) shows the AXUV array, which provided radiation signals.

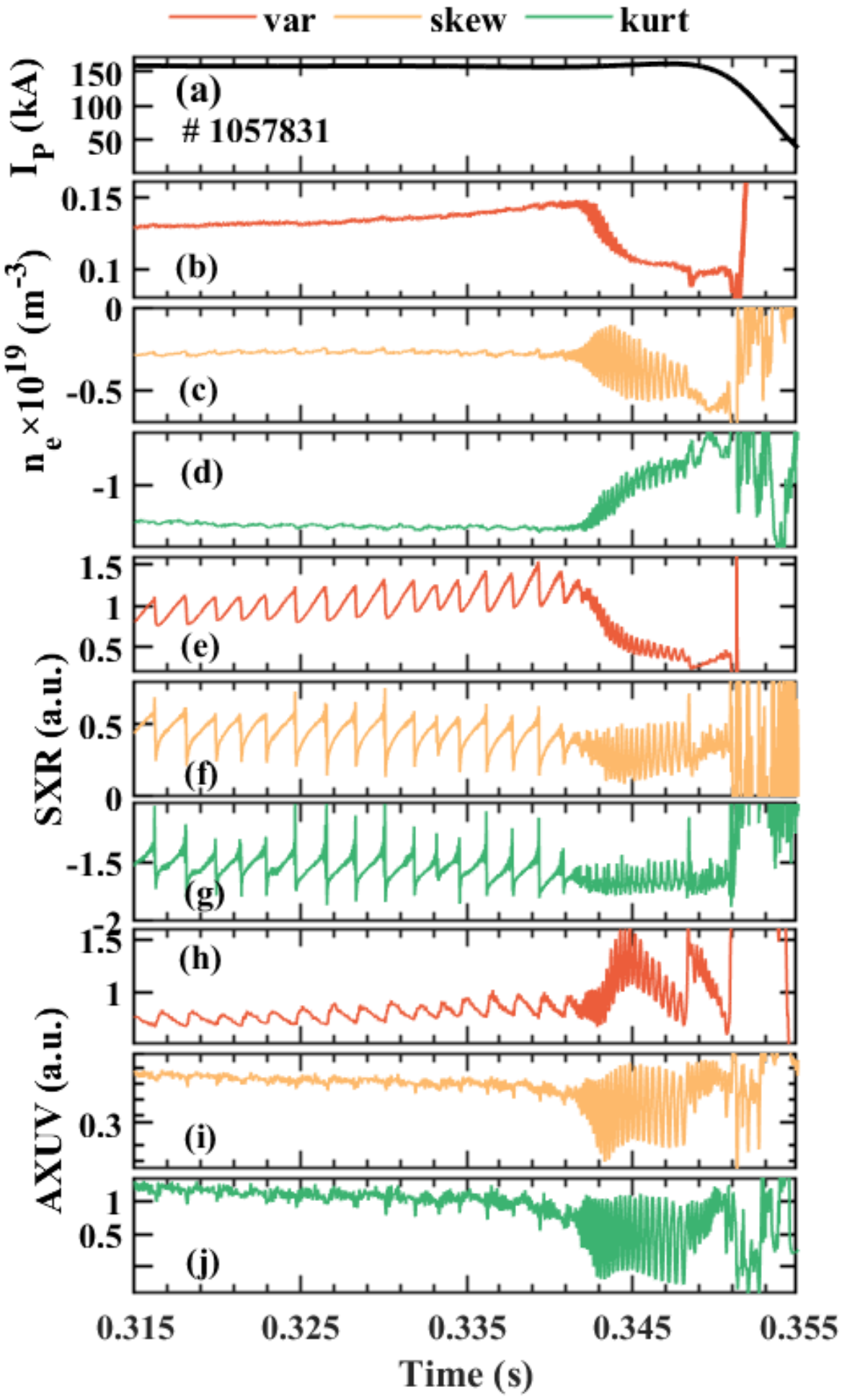

Figure 3 The disruptive discharge #1057831 range from 0.315s to 0.355s of HOS features in J-TEXT. The orange, yellow and green lines represent *var*, *skew*, and *kurt*, respectively. a) shows the plasma current of the discharge; b), c) and d) shows the FIR array, which provided density signals; e), f) and g) shows the SXR array; h), i) and j) shows the AXUV array, which provided radiation signals.

## *2.2 Disruption classifier: Dropouts meet Multiple Additive Regression Trees (DART)*

Tree-based models can be more accurate than neural networks and linear models in many applications, such as tabular-style datasets, where features are individually meaningful [50]. Features after PGFE are also individually meaningful for the disruption prediction tasks. Gradient boosting decision tree (GBDT) [51] is a tree-based algorithm that gives a prediction model in the form of an ensemble of weak prediction models, typically decision trees. Disruption prediction using LightGBM, proposed based on GBDT, was introduced in our previous work [33,34]. However, LightGBM and other GBDT-based algorithms like Multiple Additive

Regression Trees (MART) can be over-sensitive to the contributions of the few initially added trees. This is not quite good enough in interpretability research, as it may lead to the model learning wrong knowledge. Dropouts meet Multiple Additive Regression Trees (DART), which uses the trick of the dropout setting in deep neural networks [52] to drop the existing decision trees randomly. DART iteratively optimizes the boosted trees from the remaining set of decision trees could alleviate the over problem. Therefore, DART, a tree-based model, is selected as the disruption classifier of IDP-PGFE. If no tree is dropped, DART is, in fact, the traditional GBDT. If all trees are dropped, DART is equivalent to the random forest (RF) because the regression tree has to be regenerated each time, and there is no previous regression tree to base on.

### *2.3 Explainer: SHAP*

SHAP is based on the game theoretically optimal Shapley values [53], which is a method from coalitional game theory to figure out how fairly distribute the "pay-out" (prediction) among the "players" (features). Note that Shapley value is not the difference in prediction compared by removing feature(s) from the model. The Shapley value is the contribution of a feature value to the difference between the actual prediction and the mean prediction when given the current set of feature values. SHAP is also an additive feature attribution method, which uses a simpler explanation model than any interpretable approximation of the original model. Explanation models use simplified inputs $x'$ that map to the original inputs through a mapping function $x = h_x(x')$. An explanation model $g$ can be expressed as:

$$g(z') = \phi_0 + \sum_{j=1}^{M} \phi_j z_j' \tag{2-10}$$

Where $z' \in \{0,1\}^M$, $z'$ = 1 means that the corresponding feature value is "present" and 0 that it is "absent". $M$ is the number of simplified input features, and $\phi_{\mathrm{i}} \in R$ is the feature attribution for a feature $j$, the Shapley values. The Shapley values $\phi$ $(f, x)$ expressed as:

$$\phi_j(f,x) = \sum_{z' \subseteq x'} \frac{|z'|!(M-|z'|-1)!}{M!}[f_x(z') - f_x(z' \setminus j)] \tag{2-11}$$

where $|z'|$ is the number of non-zero entries in $z'$ and $z' \setminus j$ denotes setting $z'$ = 0. The Shapley values $\phi_i$ $(f, x)$, explaining a prediction $f(x)$, are an allocation of credit among the features in $x$ (the features extracted through PGFE) and are the only allocation satisfying three desirable properties. The first one is local accuracy, which ensure the explanation model $g$ at least match the predictor $f$. The second one is missingness to ensure features missing in the original input to have no impact, which means if $z'$ = 0, the importance attributed is also 0. The third one is consistency (also called monotonicity in game theory researches), which states that if a feature is more important in one model than another, the importance attributed to that feature should also be higher.

SHAP is a unified approach to interpreting model predictions, which has the advantages of a solid theoretical foundation in game theory, the only allocation of credit among the features and consistency of global and local interpretations. However, it is possible to create intentionally misleading interpretations with SHAP, which can hide biases [54], which need to be careful in explaining the predictor.

## 3. Dataset description

This section will describe the dataset selected to train, valid and test in IDP-PGFE. The electron cyclotron resonance heating (ECRH) system was successfully commissioned at the beginning of 2019 campaign[55,56]. The dataset of J-TEXT contains 1791 discharges out of 2017-2018 campaigns and 2021-2022 campaigns with the accessibility and consistency of the diagnostics channels, including discharges applied to ECRH and RMPs. All types of disruptions were included except intentional ones triggered by massive gas injection (MGI) or shattered pellet injection (SPI) and engineering tests. Both training and validation sets are selected randomly from the 2017-2018 campaigns. The test set is selected randomly from the 2017-2018 and 2021-2022 campaigns. A split of datasets is shown in table 2. 212 disruptive and 1199 non-disruptive discharges are selected as the training set. 80 disruptive and 80 non-disruptive discharges are selected as the validation set. 110 disruptive and 110 non-disruptive discharges are selected as the test set. The locked mode detectors in the 2017-2018 campaigns are ex-vessel saddle loops, while in the 2021-2022 campaigns are $B_{\mathrm{r}}$ HFS Mirnov probes [48]. The Mirnov probes in the poloidal array in the 2017-2018 campaigns are from the poloidal probe array 1, while the 2021-2022 campaigns are from the poloidal probe array 2. All discharges are split into slices from the flat-top of plasma current to current quench (CQ) time per 0.1ms, the same as the sampling rate of plasma current in J-TEXT. An automatic criterion has been applied to detect sudden large drops in $I_{\mathrm{P}}$, and marked the beginning of the drop as CQ time with 1 ms resolution. Then the results are visually checked and corrected by human experts. The phase between the CQ time and a time threshold indicates the unstable phase of each disruptive discharge. The unstable phase and time threshold of each discharge can be determined manually [22] by a statistical analysis, either equal for each discharge [10] or individually for each discharge [23,57]. An automatic approach has been used to determine the unstable phase and time threshold by finding the best performance of the model by scanning the time threshold from 5ms to 50ms in J-TEXT. The time threshold equalled to 25ms before CQ time achieved the best performance. The "unstable" slices in disruptive charges are labelled as "disruptive", and all the slices in non-disruptive discharges are labelled as "non-disruptive". The disruptive slices are totally from part of the disruptive discharges, and the non-disruptive slices are from all the non-disruptive discharges, which leads to a significant imbalance of the dataset. Therefore, we increased the weights of disruptive slices and randomly dropped a portion of non-disruptive slices to balance the two kinds of slices.

Table 2 Split of datasets of the predictor

| | Non-disruptive | Disruptive |
|---|---|---|
| No. training shots | 1199 | 212 |
| No. validation shots | 80 | 80 |
| No. test shots | 110 | 110 |

## 4. Predictive performances of IDP-PGFE

This section shows the high performance and low data required abilities of IDP-PGFE based on PGFE. Disruption prediction is a binary classification task, of which a confusion matrix is often used to evaluate performance. True positive (TP) in disruption prediction refers to a successfully predicted disruptive discharge. False positive (FP) refers to non-disruptive discharge predicted as disruptive also called false alarm. True negative (TN) refers to a non-disruptive discharge predicted as non-disruptive too. False negative (FN) refers to a disruptive discharge not predicted as disruptive. Miss alarm and tardy alarm both belong to FN. Short warning time should be reckoned as tardy alarm due to the requirement of the disruption mitigation system (DMS). In IDP-PGFE, any predicted disruption with a warning time of less than 10ms is considered FN. The evaluation indicators of a disruption predictor are the receiver operating characteristic curve (ROC), which included true positive rate (TPR), false positive rate (FPR) and area under the ROC curve (AUC). TPR and FPR are calculated as follows:

$$\mathrm{TPR} = \frac{TP}{TP + FN} \tag{4-1}$$

$$\mathrm{FPR} = \frac{FP}{FP + TN} \tag{4-2}$$

The DART will give a result between "0" ("non-disruptive") and "1" ("disruptive"), which can be binarily classified by manually setting a model threshold. Each model threshold corresponds to a set of TPR and FPR. The ROC curve is created by plotting TPR against FPR at various threshold settings.

To make a fair comparation, two benchmark disruption predictors using raw signals (DPRS) are also shown in this section. The training, valid and test sets of DPRS are the same as IDP-PGFE. One version of DPRS called DPRS-base used all the channels of diagnostic signals (92-dimensions) shown in table 1. Another version of DPRS called DPRS-light used selected channels to keep a same input dimension (28-dimensions) with the IDP-PGFE. The channels selected for DPRS-light is shown in table 3. AXUV and SXR arrays selected the channels from the chord distance about 0, ± 0.5*a* and ± *a*. FIR array selected the channels from the chord distance about 0 and ± *a*.

Table 3 Channels of diagnostics selected for DPRS-light.

| | |
|---|---|
| Channels of diagnostics | Mirnov probes in poloidal array (2) |
| | Mirnov probes in toroidal array (2) |
| | Locked mode detectors (4) |
| | AXUV array (5) |
| | CIII |
| | SXR array (5) |
| | FIR array (3) |
| | $B_t$, $I_p$ |
| | $I_{hf}$, $I_{vf}$ |
| | $d_r$,$d_z$ |

In this section, the performances of IDP-PGFE are reported in detail for J-TEXT. Section 4.1 describes the results of IDP-PGFE. As described in the introduction, PGFE has the ability on low data learning. The cases of low data learning via IDP-PGFE with fewer data are addressed in section 4.2. The hyperparameter search determines the hyperparameters of each best performance model in this section.

### *4.1 IDP-PGFE performances on J-TEXT*

In this part, IDP-PGFE was trained with full training set data. The performances of IDP-PGFE, DPRS-base and DPRS-light are shown in figure 4. The orange, navy-blue and light-blue line represents the ROC curves of IDP-PGFE, DPRS-base and DPRS-light, respectively. Although the AUC value of DPRS-base is 0.935, better than it of DPRS-light, the performance of IDP-PGFE is more outstanding. The AUC value of DPRS-light is a little smaller than DPRS-base, which indicates that only increasing the amount of input data (input dimension) is not effective in improving the performance of the disruption predictor. PGFE could significantly improve the performance of the disruption predictor by extracting disruption-related features. This leads to the conclusion that the PGFE-based IDP-PGFE is superior in predictive performance to a disruption predictor using the raw signals.

Figure 5 shows the accumulated percentage of disruption predicted versus warning time with the fixed model threshold = 0.5. Due to J-TEXT being a small-sized tokamak with a relatively smaller time scale for disruption to take place, the warning time is shorter than that for JET, EAST, DIII-D, or other large and medium sized tokamaks. Therefore, the warning time could be selected as 10ms. The electromagnetic particle injector (EPI) could react by the trigger advanced 10ms [58]. The warning time of 30ms should also ensure a considerable accumulated percentage (TPR>90%) of disruption is predicted for other mitigation methods to react. The final performance of IDP-PGFE is TPR = 97.27%, FPR = 5.45% with a tolerance of 10ms.

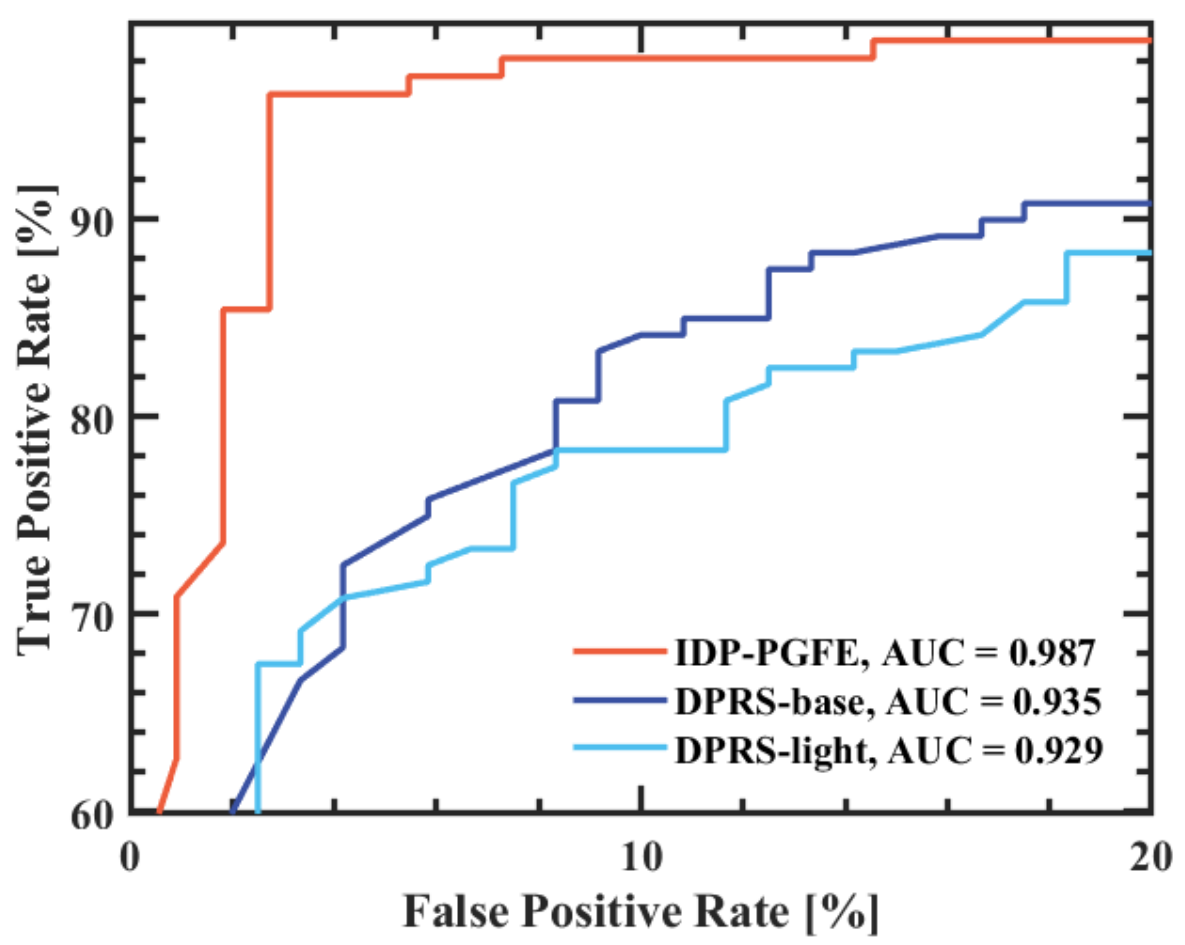

Figure 4 The ROC curves of IDP-PGFE, DPRS-base and DPRS-light. The FPR axis is from 0% to 20%. The TPR axis is from 60% to 100%. The orange, navy-blue and light-blue line represents the ROC curves of IDP-PGFE, DPRS-base and DPRS-light, respectively.

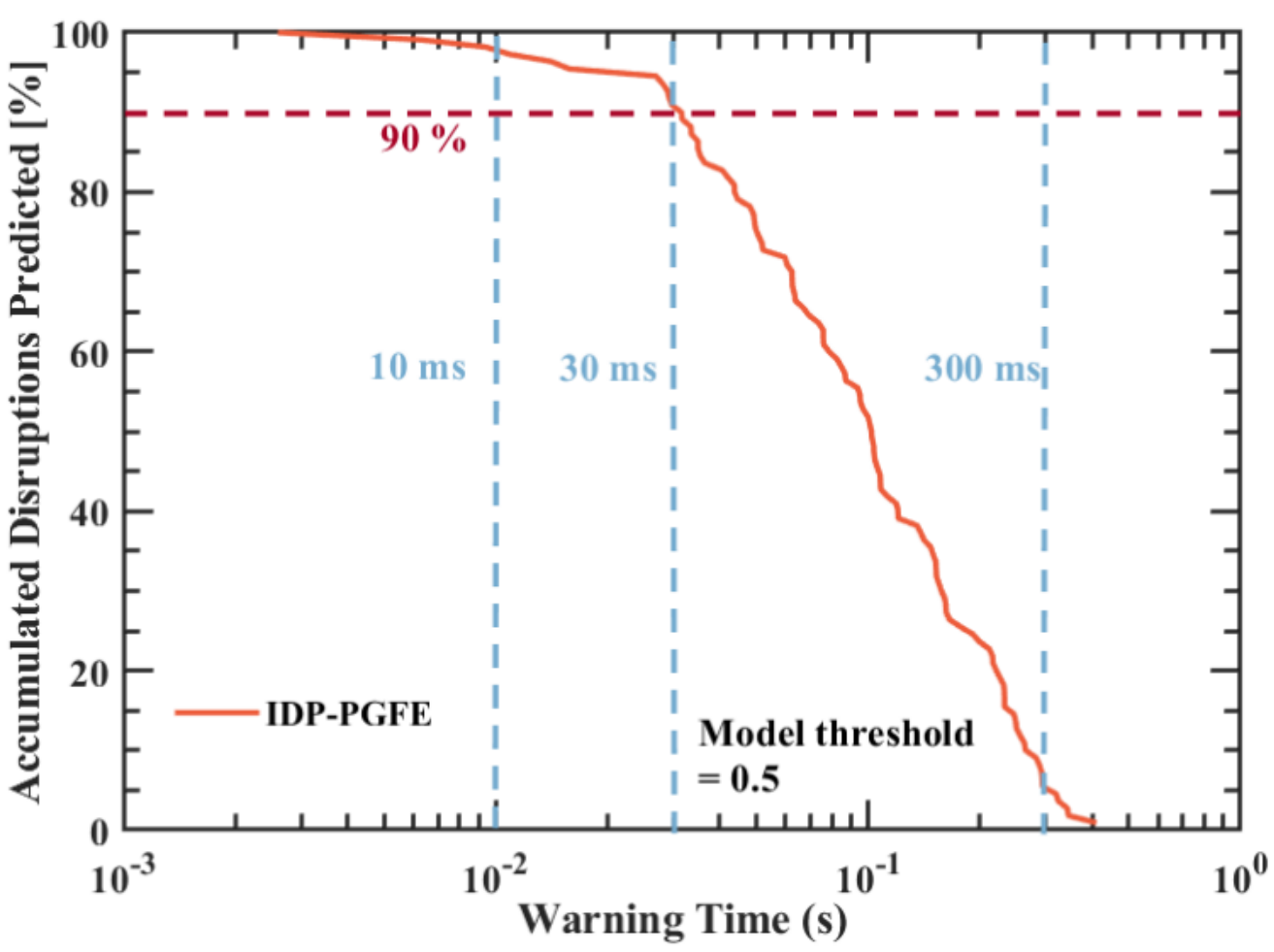

Figure 5 The accumulated percentage of disruption predicted versus warning time. The model threshold is fixed as 0.5. The red dashed line represents the accumulated percentage of disruption predicted equals to 90%. The light blue dashed lines represent the warning time of 10ms, 30ms and 300ms.

### *4.2 Low data learning via IDP-PGFE*

In this part, IDP-PGFE was trained with less data, which are 80%, 60%, 40%, 20%, 10%, 5% of the training set. The ROC curves of IDP-PGFE, DPRS-base, and DPRS-light trained by 100% (full), 60%, 40%, 10%, and 5% data sizes are shown in figure 6. As the training data size decreases, the performance of IDP-PGFE degrades. The performance of IDP-PGFE with a 10% data size reaches that of DPRS-base and is higher than that of DPRS-light. However, the performance of IDP-PGFE with a 5% data size degrades significantly. IDP-PGFE with 10% data size is still acceptable as a disruption predictor (AUC = 0.939, the best TPR = 89.09%, FPR = 10%), whereas with 5% data size is not (AUC = 0.845, the best TPR = 75.45%, FPR = 15.45%).

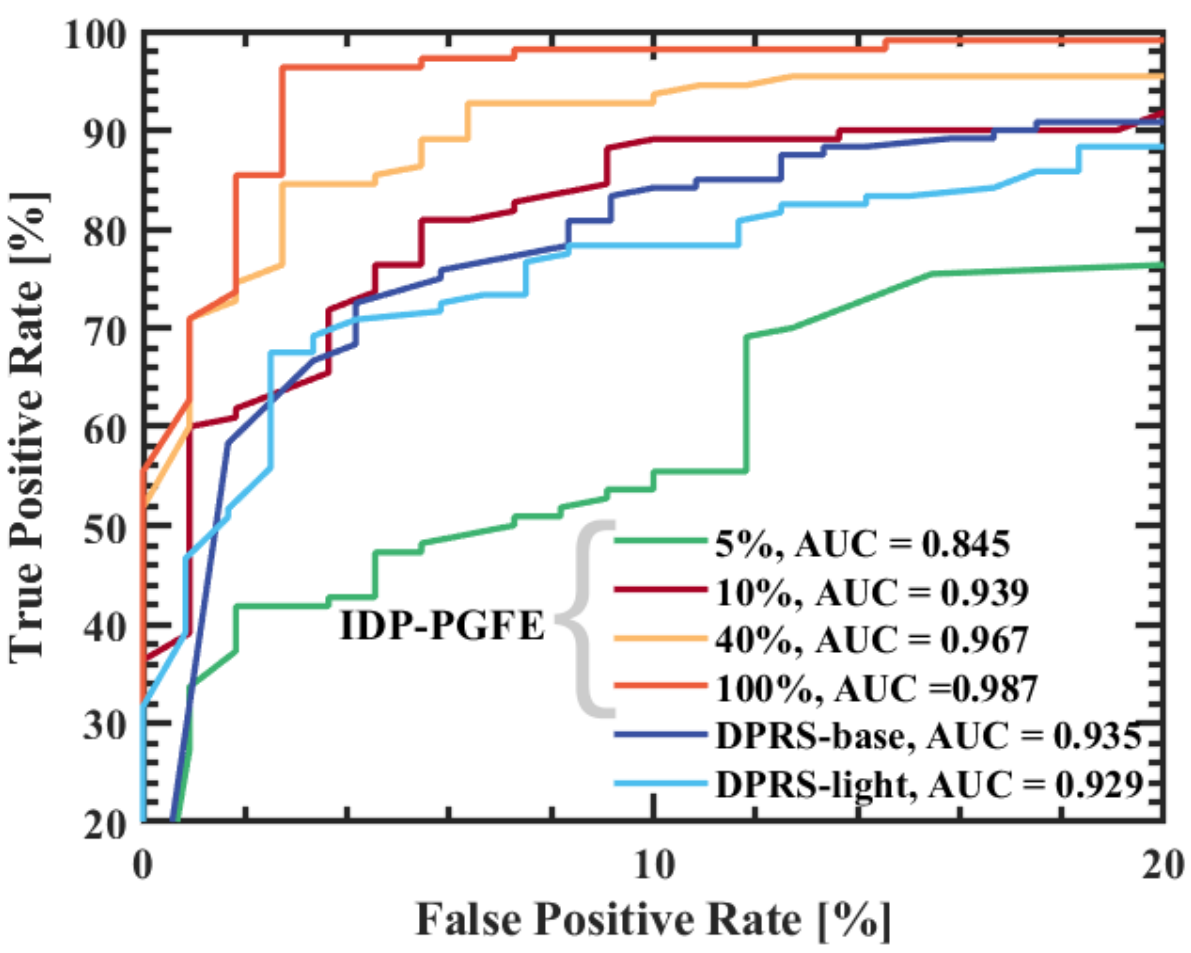

Figure 6 The ROC curves of IDP-PGFE trained by different data sizes, DPRS-base, and DPRS-light trained by full data sizes. The FPR axis is from 0% to 20%. The TPR axis is from 20% to 100%. Five coloured lines represent five kinds of data size, respectively (100% – orange, 40% – yellow, 10% – deep-red, and 5% – green). The navy-blue and the light-blue lines represent the DPRS-base and DPRS-light ROC curves. The AUC value of IDP-PGFE with the 10% data size is the same as the AUC value of DPRS-light and a little bit smaller than DPRS-base. The AUC value of IDP-PGFE with the 5% data size is the smallest.

The low data learning ability does not benefit from the disruption classifier DART but from the feature extractor PGFE. DPRS-base and DPRS-light were also trained with fewer data as the benchmarks. The evolution of the AUC values with the training data size is shown in figure 7. The AUC value of DPRS-base and DPRS-light are similar to each other with the data size of 80%, 60%, 40% and 20%. When the training data size smaller than 20% the AUC value of DPRS-light degrades faster than DPRS-base. As the training data size decreases, the performance of DPRS-base and DPRS-light both degrades faster than IDP-PGFE. The performance of IDP-PGFE and DPRSs has a significant reduction at 10% data size and 20% data size, respectively. It is worth noting that the 10% data size only includes about 20 disruptive discharges and 120 non-disruptive discharges, which are quite a small data requirement for the disruption prediction task. PGFE as a feature extractor could obviously reduce the data requirement of the disruption predictor by the feature extraction specific to tokamak disruption.

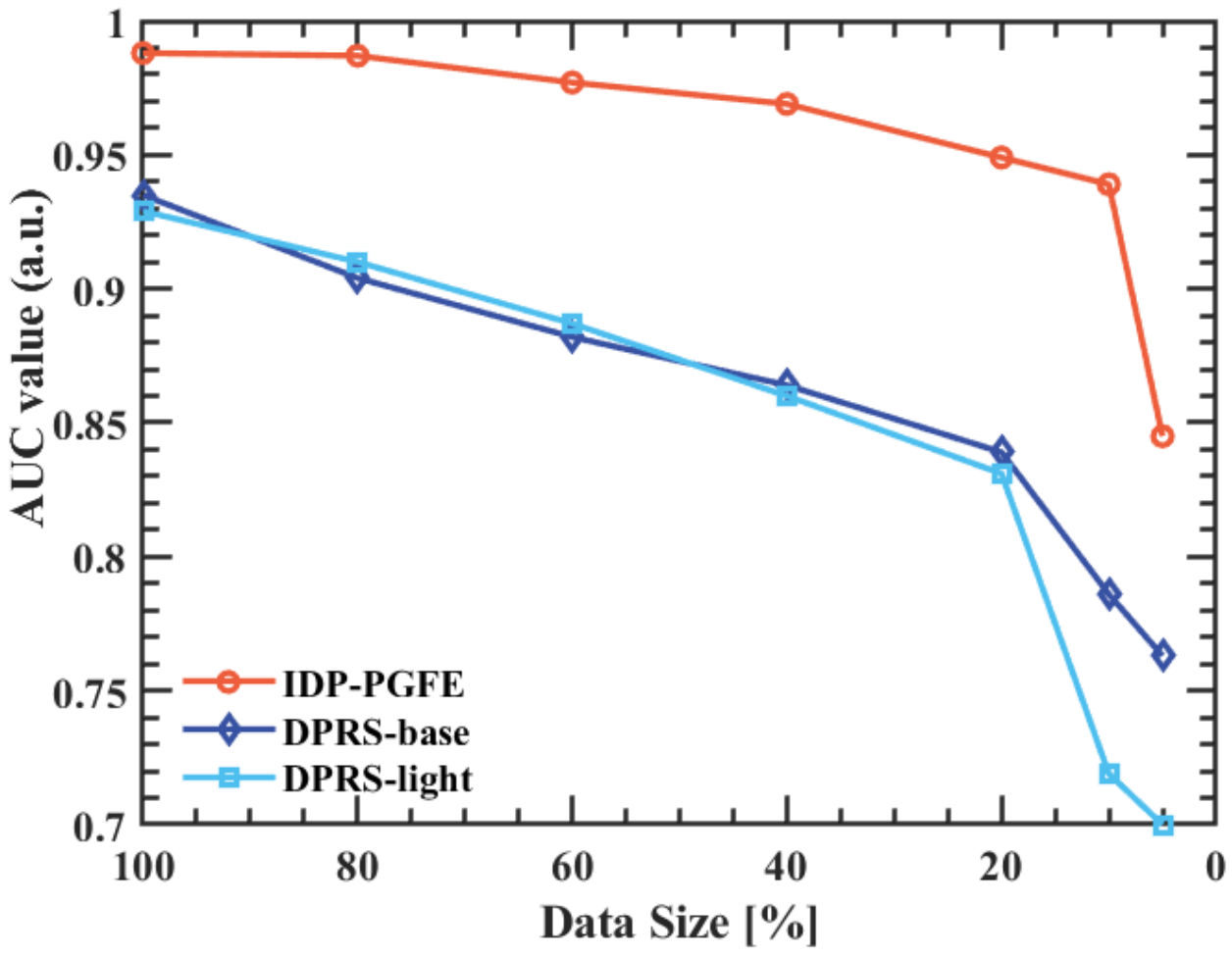

Figure 7 The AUC value versus data size of the training set for two kinds of models. The orange, navy-blue and light-blue line represents IDP-PGFE, DPRS-base and DPRS-light, respectively. The x-axis decreases from 100% to 5%. The AUC value of DPRS-base and DPRS-light decreases faster than that of IDP-PGFE as the data size decreases. The AUC value of IDP-PGFE ramps down from 0.939 to 0.845 with the data size from 10% to 5%, while the AUC value of DPRS-base ramps down from 0.839 to 0.763 and the AUC value of DPRS-light ramps down from 0.831 to 0.7 with the data size from 20% to 5%.

## 5. Interpretability study and the application on J-TEXT experiment

This section will describe the interpretability study by IDP-PGFE and its application on J-TEXT disruption experiments. Benefit from PGFE to extract meaningful features, IDP-PGFE will be more interpretable than using the raw diagnostic signals. The interpretability application presented is not aiming real-time disruption prediction but gives a possible mean of inferring the underlying cause of the disruption and how the interventions affect the disruption process. IDP-PGFE can also be used in real-time as a guide for disruption avoidance, which is beyond the scope of this article. Sections 5.1 will show the global interpretability of IDP-PGFE and section 5.2 will describe the RMP experiment applications.

### *5.1 Global interpretability study of IDP-PGFE*

The outstanding performance of IDP-PGFE can somewhat reflect that IDP-PGFE has learned enough correct knowledge related to J-TEXT disruption. Otherwise, it would not be possible to have such high accuracy. Moreover, checking the model interpretation results with widely accepted disruption precursors can also act as a test to verify that the model has learned physically meaningful patterns instead of some leaked information in the data. Nevertheless, understanding how IDP-PGFE recognizes "disruptive" and "non-disruptive" is more important for inferring the underlying cause of the disruption. SHAP is also an attribution technique that could not only provide the contributions of features but also how feature value impacts model output. We will still study the interpretability of IDP-PGFE from the four categories of the PGFE, MHD instabilities, radiation, density related features and basic PCS signals. Figure 8 shows the SHAP value of different features and their relations with feature value.

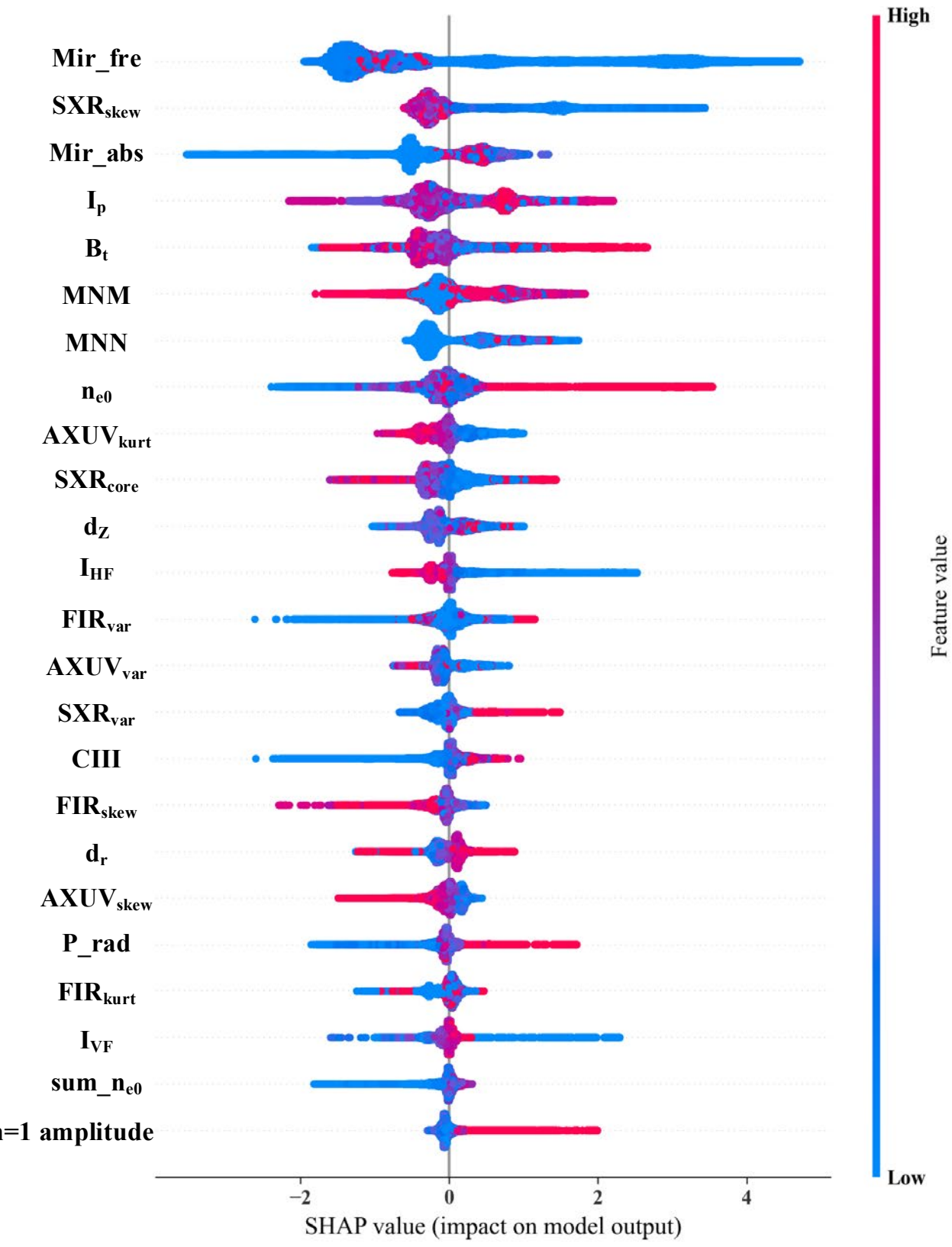

Figure 8 The SHAP value of different features and their relations with feature value. The width of bar in the SHAP result represent the number of the samples. The larger the width of the bar, the larger the number of samples. The order of the features represents the contributions of features. The colormap represents the feature value of each feature, red means high and blue means low. A positive SHAP value represents a "disruptive" impact on the model, while a negative SHAP value represents a "non-disruptive" impact on the model.

***MHD instabilities related features contributions*:** The frequency of Mirnov probe *Mir_fre* reflected the frequency of MHD instabilities is the most contributed feature. The lower value of *Mir_fre* may both contribute to "disruptive" and "non-disruptive". If the frequency of the Mirnov probe slowdown, the contribution is "disruptive", which reflects the locked mode process. Otherwise, if the frequency of the Mirnov probe is always about 0, the contribution is "non-disruptive", reflecting no MHD instabilities. The higher value of *Mir_fre* contributes more to "non-disruptive", which means the high-speed rotating mode is not the direct reason for disruption. The two cases can be distinguished by the 3rd ranked contribution feature, the intensity of Mirnov probe *Mir_abs*. The lower value of *Mir_abs* contributes more to "non-disruptive", which represents the no MHD instabilities case. J-TEXT is a small-sized tokamak with a relatively smaller time scale for LM caused disruption. Therefore, *n = 1 amplitude* is not an effective time-sensitive disruption precursor. The typical discharge shown in section 2.1.1 also proves that the time between the LM and disruption is short. The *n = 1 amplitude* is the 24th ranked contribution feature, and the higher value contributes more "disruptive". *MNM* is the 6th ranked contribution feature and *MNN* is the 7th ranked contribution feature. The higher value of *MNM* ($m \geq 2$) both contribute more to "disruptive" and "non-disruptive". The case that contributes more to "disruptive" may relate to growth of $m/n = 2/1$ (TM) and multi-magnetic island overlaps. In contrast, the other case may relate to edge magnetic island, which could even improve the plasma confinement [59].

***Radiation related features contributions*:** The 2nd ranked contribution feature is $SXR_{skew}$, and the lower value of $SXR_{skew}$ contributes more to "disruptive". The lower value of $SXR_{skew}$ does not mean that the soft *x*-ray radiation is high. Figure 9 shows the SHAP value of $SXR_{skew}$ with respect to the value of $SXR_{skew}$ and $SXR_{core}$. This figure shows that only lower $SXR_{core}$ and lower $SXR_{skew}$ together contributed more to "disruptive", reflecting the general decrease of soft *x*-ray radiation. The 15th ranked contribution feature is $SXR_{var}$, and the higher value of $SXR_{var}$ represents the more peaked profile, which contributes more to “disruptive”. This is usually a precursor to the density limit disruption. The 9th ranked contribution feature is $AXUV_{kurt}$, and the lower value of $AXUV_{kurt}$ represents the thinner profile, which contributes more to "disruptive". The thinner profile usually represents more significant gradients of the profile. The 20th ranked contribution feature is *P_rad*, and the higher value of *P_rad* represents the more losses of plasma radiation energy, which contributes more to “disruptive”.

***Density related features contributions*:** The density related features are intuitive to understand. The most relevant feature among the density related features is $n_{e0}$, the 8th ranked contribution feature. The higher value of $n_{e0}$ represents the density limit disruption. No matter how $FIR_{var}$, $FIR_{skew}$ and $FIR_{kurt}$ contribute, J-TEXT cannot escape the density limit disruption after all. The higher value of $FIR_{var}$ and lower value of $FIR_{skew}$ contribute more to "disruptive". The could provide more detailed information on the density limit disruption, which is the more peaked density profile and the more general increase of plasma density.

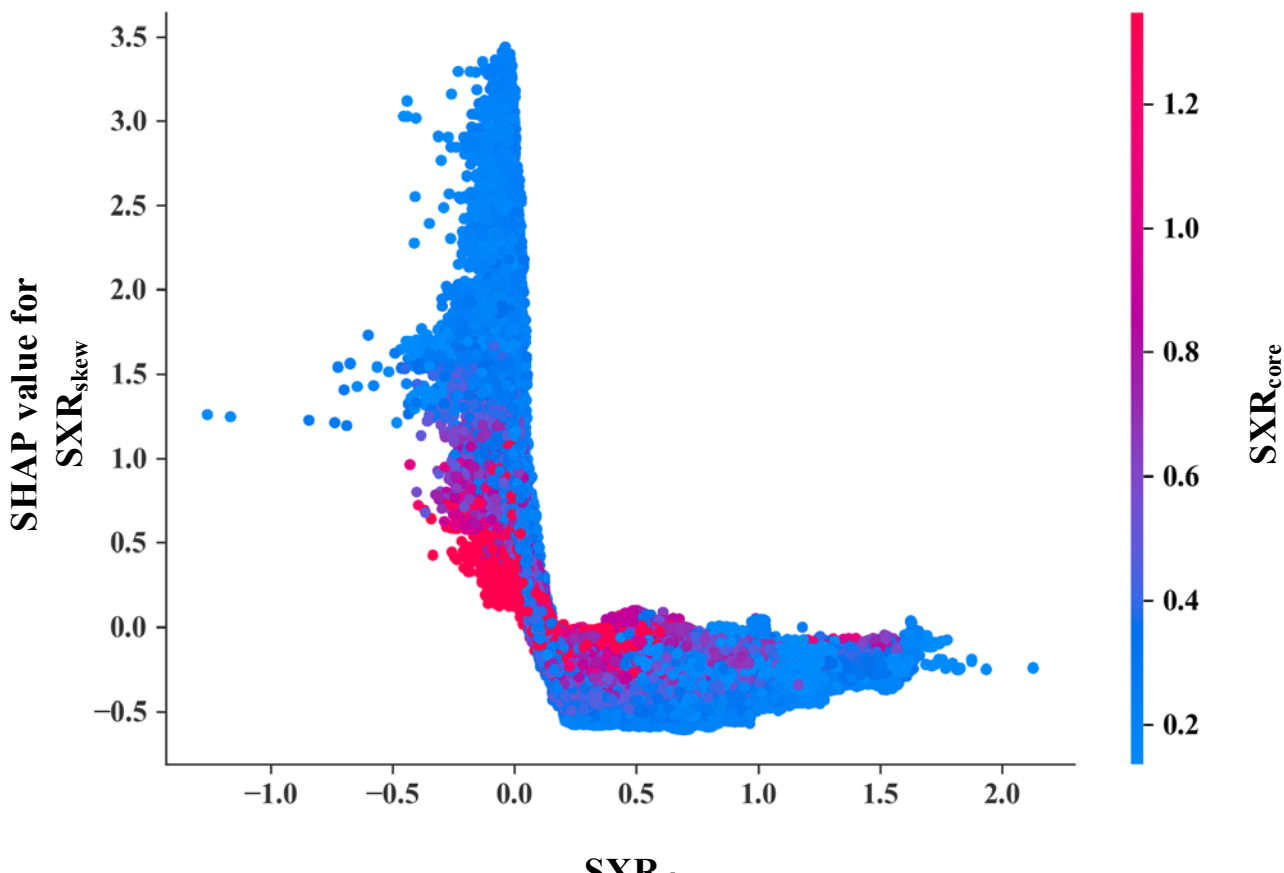

Figure 9 The SHAP value of $SXR_{skew}$ with respect to the value of $SXR_{skew}$ and $SXR_{core}$. The x-axis represents the value of $SXR_{skew}$; the colormap represents the value of $SXR_{core}$; the y-axis represents the SHAP value of $SXR_{skew}$.

***Basic PCS signals contributions:*** Notably, $I_P$ and $B_t$ get the 4th and 5th ranked contribution features, respectively. Even if $I_P$ and $B_t$ do not change much in one discharge, their contribution is still significant because they are basic parameters of plasma. The value of $I_P$ and $B_t$ along has no impact on "disruptive" or "non-disruptive". Both $d_Z$ and $I_{HF}$ could reflect the vertical displacement. The $I_{HF}$ should be larger than 0 (based on the value of TF) when the plasma is stable (no vertical displacement) in J-TEXT. Higher $I_{HF}$ will raise the plasma and vice versa. Therefore, the higher or lower $I_{HF}$ indicates that the vertical displacement tends to go down or up. According to the SHAP result, the lower $I_{HF}$ is more contributed to "disruptive" reflecting that the vertical displacement goes up usually causes disruption in J-TEXT. As shown in figure 10, this proves that the plasma usually tends to drift up when approaching disruption. So $I_{HF}$ needs to pull the plasma back. We have yet to find the reasons, which may be due to the electromagnetic force on J-TEXT or the position of the gas puffing fuelling. It requires further investigation and analysis in the future. Both $d_r$ and $I_{VF}$ could reflect the horizontal displacement. The higher value of $d_r$ both contributes more to "disruptive" and "non-disruptive" may represent that in J-TEXT, the horizontal displacement to LFS limiter is salvageable.

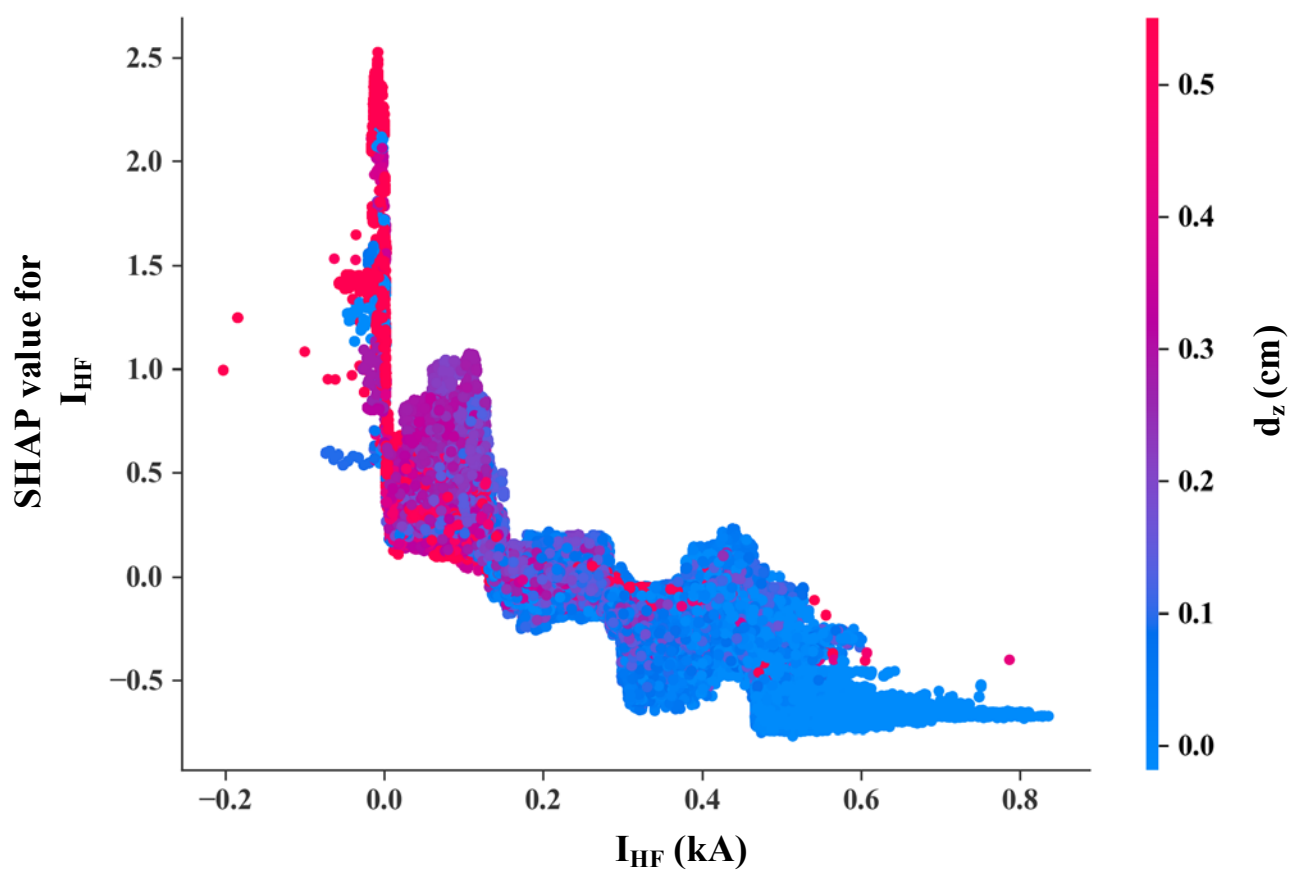

Figure 10 The SHAP value of $I_{HF}$ with respect to the value of $I_{HF}$ and $d_z$. The x-axis represents the value of $I_{HF}$; the colormap represents the value of $d_z$; the y-axis represents the SHAP value of $I_{HF}$.

### *5.2 Interpretability application on RMP experiment*

The analysis above is only a preliminary study on the interpretability of IDP-PGFE. Much of the analysis is speculative and not conclusive. Even though the study is preliminary, it still fits the experimental phenomenon well and helps experimenter to find possible experimental analysis direction for further analysis in J-TEXT experiments.

The experiments investigating the influence of RMPs on disruption due to continuously increasing density towards density limit in J-TEXT were carried out in the 2022 spring. In the RMP experiments ($I_p$ = 120 kA, $B_t$ = 2.1 T, $I_c$ = 0-4 kA) it is found that the application of RMPs raised the density limit at disruption. Two typical discharges in this experiment have been selected in the test set. The overviews of the experiment are shown in figure 11. The discharge without the application of RMP also reached the density limit of about 0.79$n_G$. In the RMP experiment, the discharges with the application of 2 kA RMPs reached the highest density limit of about 0.91$n_G$, which is a really high density limit in J-TEXT. The #1080564 discharge is a typical high $q_a$ density limit disruption discharge in J-TEXT. In this kind of discharge, multifaceted asymmetric radiation from the edge (MARFE) and detached plasma reflected in the ratio evolution of the line integral density at HFS to that at LFS [61,62]. As the density continued to rise, the ratio began to increase along with the onset of MARFE, then decreased when detached plasma happened. The application of RMPs delays the decrease of the ratio, as a result in raising the density limit and delaying the disruption in #1080550 discharge. The interpretability study of IDP-PGFE could help physicists confirm if the application of RMPs raised the density limit and find out the possible reason.

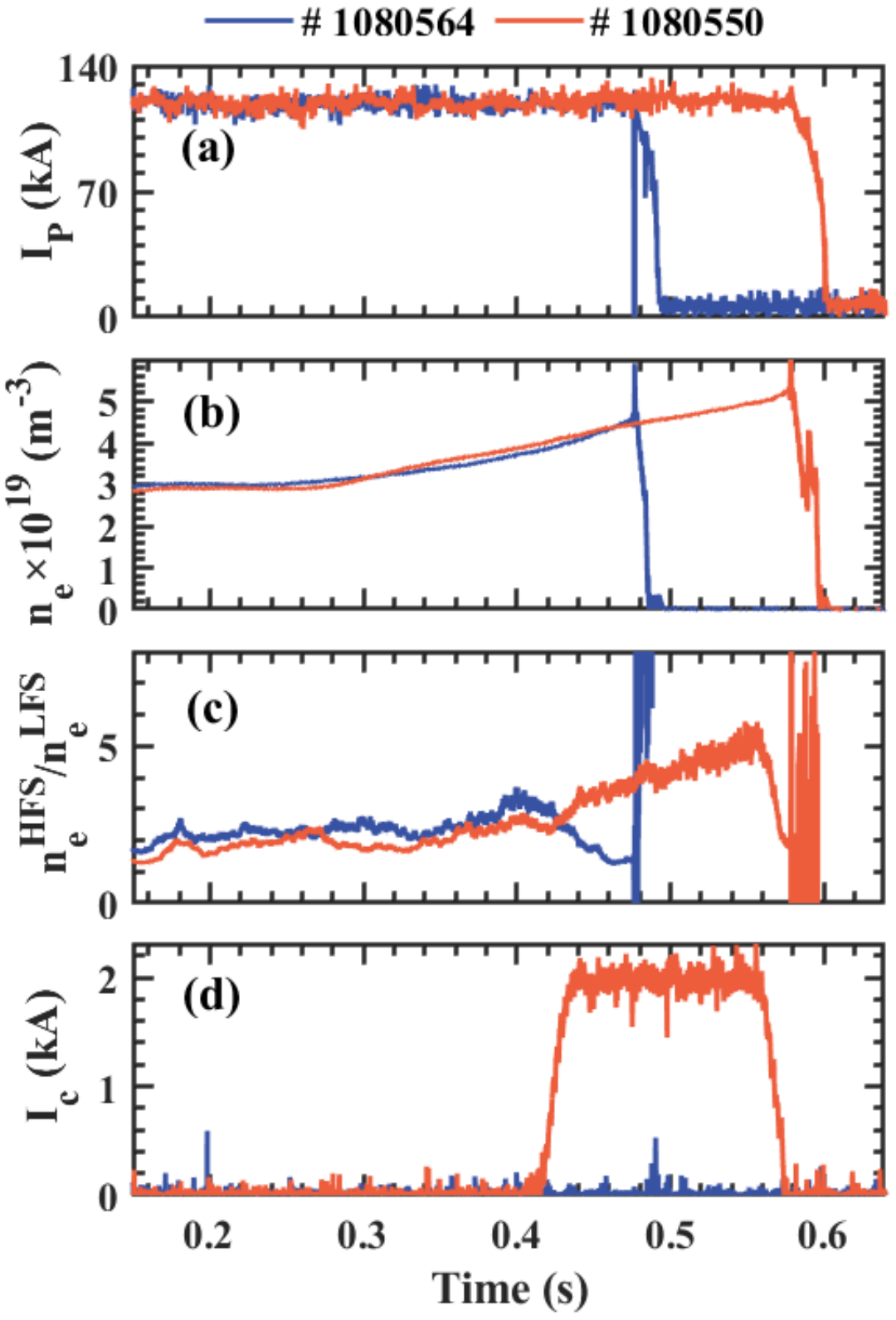

Figure 11 The overview of the two typical discharges in J-TEXT disruption experiment under RMPs. The blue line represents the discharge without the application of RMPs with the density limit of about 0.79$n_G$. The orange line represents the discharge with the application of 2 kA RMPs with the density limit about of 0.91$n_G$. a) shows the plasma current of three discharges; b) shows the line average density of middle channel; c) shows the ratio of the line integral density at HFS to that at LFS; d) shows the RMPs current. Disruption occurred after the RMPs cut off.

Figure 12 and figure 13 show the predicted result, SHAP value and four typical features in #1080564 (w/o RMPs) and #1080550 (with RMPs) discharges, respectively. The disruption time is much later than the predicted time contributed by the core density with the application of RMPs (# 1080550). This indicate that the application of RMPs delay the density limit disruption and raise the density limit.

The #1080564 discharge (w/o RMPs) is a high $q_a$ density limit disruption discharge different from #1074378. The predicted result increased because of the increase in SHAP value of *CIII*, $AXUV_{kurt}$ and *MNM*. The SHAP value of $n_{e0}$ increased only close to disruption. The SHAP value of *CIII* increased at about 0.39s (during MARFE), contributing to the predicted result of about 0.4. At about 0.42s, detached plasma happened. The SHAP value of $AXUV_{kurt}$ increased with $AXUV_{kurt}$'s decrease makes the predicted result increase. At about 0.44s, the SHAP value of *MNM* increased and contributed to the predicted result larger than 0.5. The interpretability study in the high $q_a$ density limit disruption discharge shows that before the density limit disruption, detached plasma and MHD instabilities could be the reason for density limit disruption [61].

In #1080550 discharge (with RMPs), The SHAP value of *CIII* still increased at about 0.39s (during MARFE). Because of the application RMPs, *CIII* continuing increase to about 0.56s with no detached plasma. Therefore, the value of $AXUV_{kurt}$ did not decrease, and the SHAP value of $AXUV_{kurt}$ did not increase. The most significant contribution to the predicted result is $n_{e0}$ because, in the view of IDP-PGFE, this discharge reached the density limit. After RMPs cut off, at about 0.56s, the value of $AXUV_{kurt}$ decreased and *MNM* increased, so the detached plasma and MHD instabilities appeared again and caused the disruption.

RMPs not only has an impact on MHD instabilities as investigated before [63] but also has an impact on radiation profile ($AXUV_{kurt}$). The contribution of $AXUV_{kurt}$ and *MNM* are more important than the contribution of $n_{e0}$ for underlying cause of the density limit disruption. It brings a new sight for the RMP experiment on the density limit disruption, which could guide physicists to figure out how the RMPs affect the density limit disruption process. The application of RMPs verify that the radiation (profile) and MHD instabilities has a strong connection to density limit disruption. In density limit disruption, the contribution of radiation (profile) and MHD instabilities also rises before disruption. The application of RMPs makes the contribution of radiation (profile) and MHD instabilities rise after the density limit (~0.79$n_G$), so that the density limit has been raised (~0.91$n_G$).

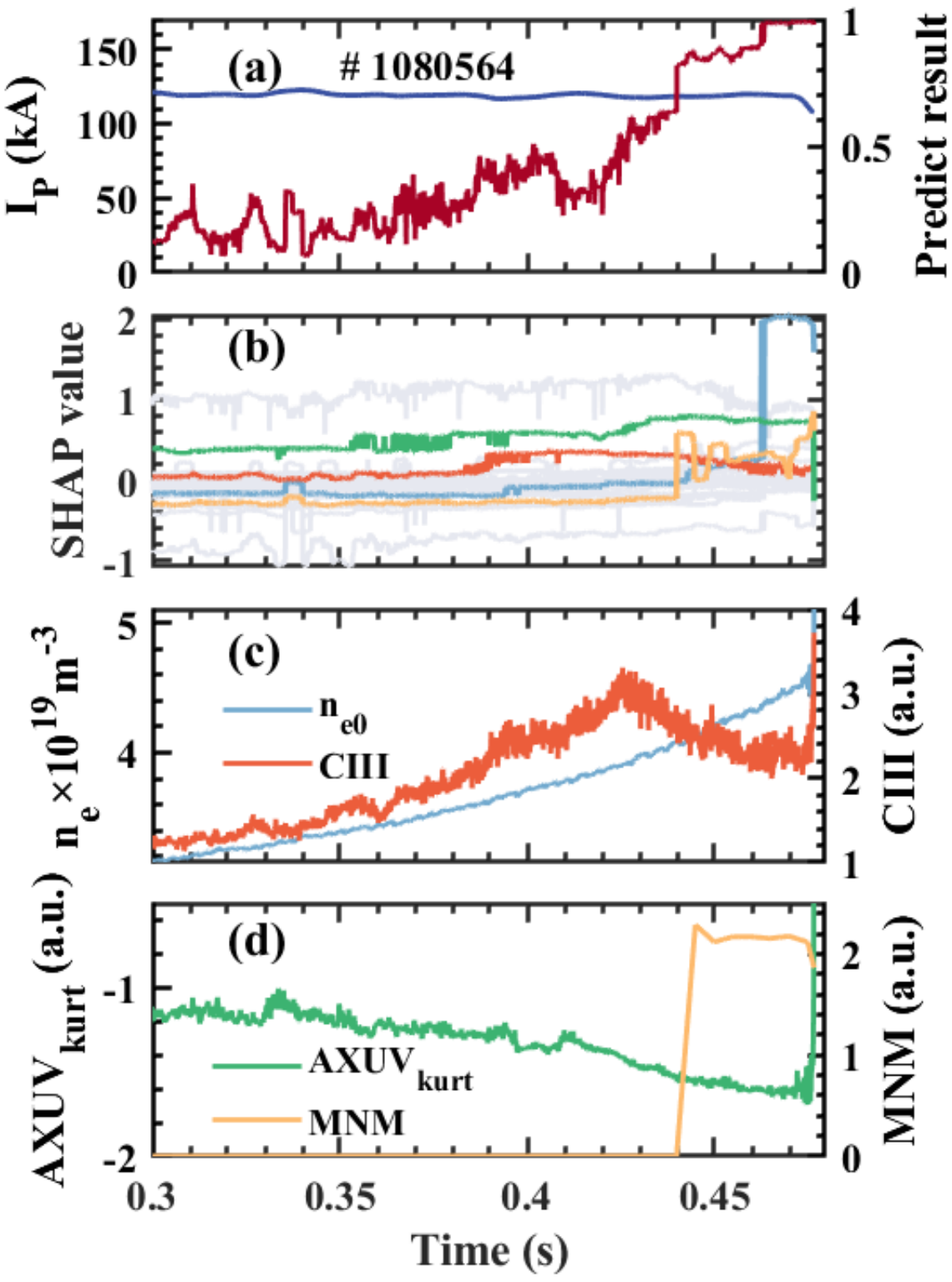

Figure 12 The predicted result, SHAP value and four typical features in # 1080564 discharge without the application of RMPs. a) shows the plasma current and predicted result; b) shows the SHAP value of all the features, the colored lines represent the four typical features; c) shows two typical features, line average density of the middle channel $n_{e0}$ (light blue) and the impurities radiation *CIII* (orange); d) shows two typical features, the kurtosis of AXUV array $AXUV_{kurt}$ (green) and the poloidal mode number *MNM* (yellow).

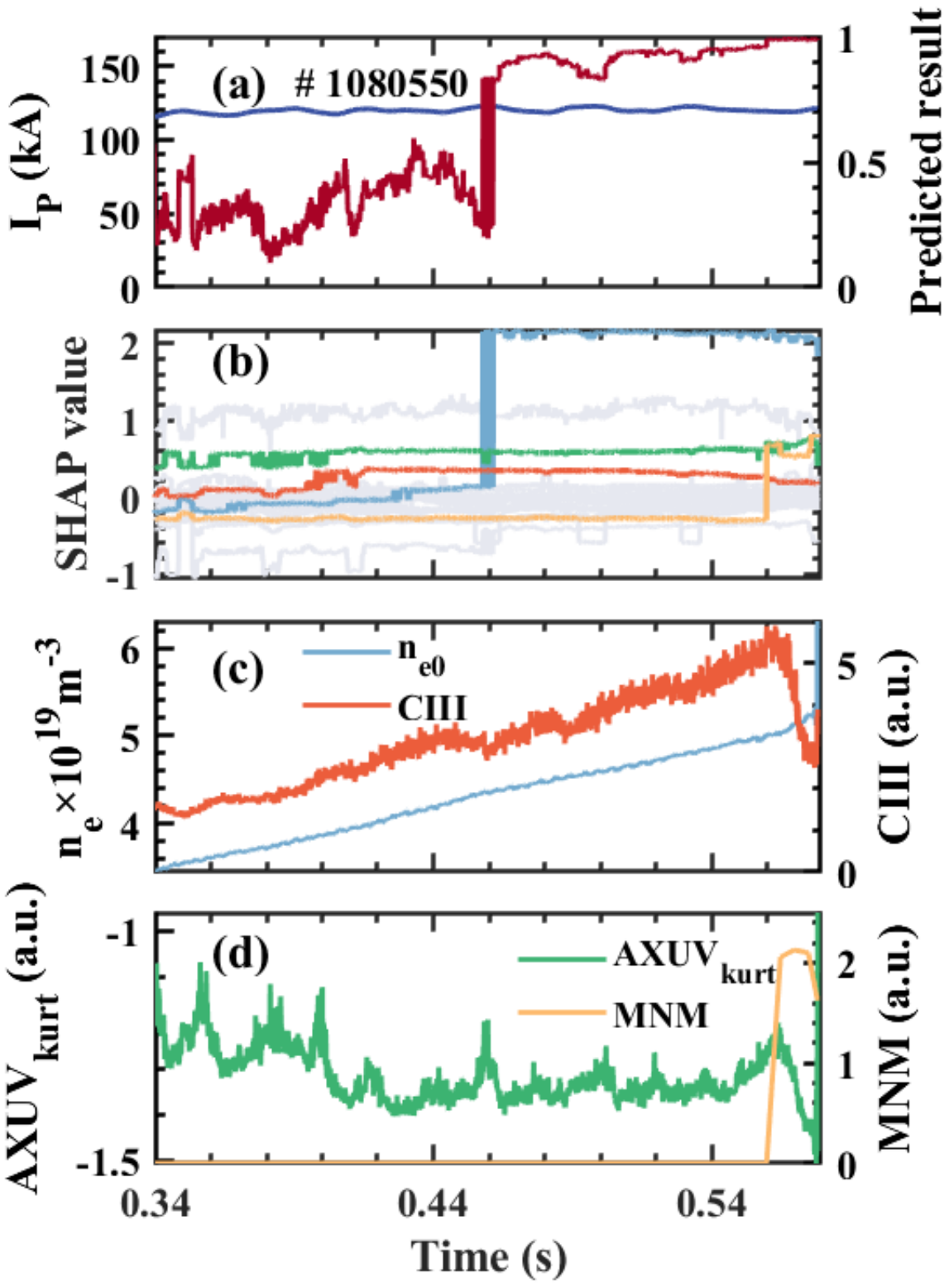

Figure 13 The predicted result, SHAP value and four typical features in # 1080550 discharge with the application of RMPs. a), b), c) and d) show the same as figure 12.

Therefore, applying RMPs can raise the density limit by delaying detached plasma and MHD instabilities before the density limit disruption. It is found that the application of

## 6. Discussion

This section will discuss the potential and limitations of IDP-PGFE. We will discuss three topics: interpretability, transferability and feasibility of real-time.

**Interpretability:** In IDP-PGFE, PGFE as a feature extractor extracts the disruption-related features. As a disruption classifier, DART is the machine learning tool to discover potential patterns and relations between features. SHAP is the attribution technique as an explainer to understand the potential patterns and relations between features. Benefit from PGFE to extract meaningful features, IDP-PGFE will be more interpretable than using the raw diagnostic signals. It means that IDP-PGFE has the potential to give a possible mean of inferring the underlying cause of the disruption and propose new conjectures. The application of RMPs affect not only the MHD instabilities but also the radiation profile, which delays the density limit disruption. More new conjectures and physical mechanisms of disruption require in-depth collaboration with physicists in the future.

However, IDP-PGFE is only applied in J-TEXT, therefore, IDP-PGFE could only learn about the disruption existing on J-TEXT. In the future, IDP-PGFE could be a tool to explore disruption event chain and disruption mechanisms by applying to dataset from different devices. Moreover, the feature extraction techniques of PGFE are not ultimate. Inverted plasma profiles, more physics features and other inductive biases of disruption could make PGFE more interpretable.

**Transferability:** IDP-PGFE has an advantage in the low data learning. Therefore, IDP-PGFE could directly train on the target dataset. However, even if the training set only required 10%, it still included about 20 disruptive and 120 non-disruptive discharges. Using domain adaptation on the features will boost the transferability of IDP-PGFE. PGFE relies on the expert experiences summarized from the dataset and the disruption-related physics. The experiences from the disruption-related physics could transfer to future devices. Moreover, the dataset to summarize the expert experiences does not necessarily have to be the training data. The expert experiences could be summarized from the dataset in the low-performance scenario or existing devices for ITER or future devices. However, PGFE could not handle future devices' new physics and device-specific features.

**Feasibility of real-time:** IDP-PGFE is based on tree models, an advantage in real-time disruption prediction. Meanwhile, PGFE makes the diagnostic signals into diagnostic-independent features, which means the lacked diagnostic signals are alternative to IDP-PGFE. The real-time tree-based disruption prediction has been tested on J-TEXT. The calculation speed can meet the requirement of the DMS. IDP-PGFE also has the potential for disruption avoidance and prevention via real-time SHAP value. The advantage of SHAP value is that we not only know the contribution of features, but also know how features change will affect the contribution. To avoid disruption, PCS only need to send different commands to auxiliary systems (like RMPs or ECRH) for different disruption causes (SHAP value).

## 7. Summary

This paper introduced an interpretable disruption predictor based on physics-guided feature extraction (PGFE) called IDP-PGFE. Actually, IDP-PGFE is a general method applicable to all tokamaks. IDP-PGFE comprises a feature extractor, a disruption classifier and an explainer based on PGFE, DART and SHAP, respectively. PGFE extracts feature with the inductive bias related to MHD instabilities, radiation and density. The feature extraction technique for PGFE might be different on different tokamaks. J-TEXT feature extractor may not work directly on other devices. But with knowledge of machine operation and diagnostic design, it can be transfer to other machines easily.

Therefore, IDP-PGFE has been first applied on J-TEXT. IDP-PGFE reaches the best performance in J-TEXT disruption prediction task benefit from PGFE. The performance of IDP-PGFE is TPR = 97.27%, FPR = 5.45% with a warning time of 10ms. PGFE could also reduce the data requirement of IDP-PGFE. The performance of IDP-PGFE using PGFE with only a 10% data size of the training dataset is similar to the performance of disruption predictor using raw signals with a full training dataset. The analysis of the contribution of features given by SHAP is not only generally consistent with existing comprehension but also helps understand some of the J-TEXT disruption processes. The interpretability application provides the analysis and brings new sight of the existing experiment phenomena. It gives a possible mean of inferring the underlying cause of the disruption and how the interventions affect the disruption process. The interpretation results and the experimental phenomenon have a high degree of conformity. The interpretability study of IDP-PGFE also gives a possible experimental analysis direction that RMPs affect not only the MHD instabilities but also the radiation profile, which delays the density limit disruption.

Last but not least, we briefly discuss three topics to discuss the potentials and limitations of IDP-PGFE. As discussed in section 6, IDP-PGFE has three potentials. First, IDP-PGFE could propose new conjectures of disruption, particularly after applying to other large and advanced tokamaks. Second, IDP-PGFE could transfer to other tokamaks with low data learning and domain adaptation. Third, IDP-PGFE could enable real-time disruption avoidance by identifying the different disruption causes.

## Acknowledgement

This work was supported by National Key R&D Program of China under Grant (No. 2018YFE0310300, No. 2022YFE03040004 and No. 2019YFE03010004) and by National Natural Science Foundation of China (NSFC) under Project Numbers Grant (No. 12075096 and No. 51821005).

## References

1. Hender, T. C. *et al.* Chapter 3: MHD stability, operational limits and disruptions. *Nucl. Fusion* **47**, S128–S202 (2007).

2. Boozer, A. H. Theory of tokamak disruptions. *Physics of Plasmas* **19**, 058101 (2012).
3. Sugihara, M. *et al.* Disruption scenarios, their mitigation and operation window in ITER. *Nucl. Fusion* **47**, 337–352 (2007).
4. de Vries, P. C. *et al.* Requirements for Triggering the ITER Disruption Mitigation System. *Fusion Science and Technology* **69**, 471–484 (2016).
5. de Vries, P. C. *et al.* Survey of disruption causes at JET. *Nucl. Fusion* **51**, 053018 (2011).
6. Berkery, J. W., Sabbagh, S. A., Riquezes, J. D., Gerhardt, S. P. & Myers, C. E. Characterization and forecasting of global and tearing mode stability for tokamak disruption avoidance. 4.
7. Berkery, J. W., Sabbagh, S. A., Bell, R. E., Gerhardt, S. P. & LeBlanc, B. P. A reduced resistive wall mode kinetic stability model for disruption forecasting. *Physics of Plasmas* **24**, 056103 (2017).
8. Sabbagh S.A. et al 2018 Disruption event characterization and forecasting in tokamaks Preprint: 2018 IAEA Fusion Energy Conf. (Gandhinagar, India, 22–27 October 2018) [EX/P6-26] (https://conferences.iaea.org/event/151/contributions/5924/).
9. Vega, J. *et al.* Results of the JET real-time disruption predictor in the ITER-like wall campaigns. *Fusion Engineering and Design* **88**, 1228–1231 (2013).
10. Rea, C. *et al.* Disruption prediction investigations using Machine Learning tools on DIII-D and Alcator C-Mod. *Plasma Phys. Control. Fusion* **60**, 084004 (2018).
11. Zheng, W. *et al.* Hybrid neural network for density limit disruption prediction and avoidance on J-TEXT tokamak. *Nucl. Fusion* **58**, 056016 (2018).
12. Yang, Z. *et al.* A disruption predictor based on a 1.5-dimensional convolutional neural network in HL-2A. *Nucl. Fusion* **60**, 016017 (2020).
13. Cannas, B. *et al.* Overview of manifold learning techniques for the investigation of disruptions on JET. *Plasma Phys. Control. Fusion* **56**, 114005 (2014).
14. Piccione, A., Berkery, J. W., Sabbagh, S. A. & Andreopoulos, Y. Physics-guided machine learning approaches to predict the ideal stability properties of fusion plasmas. *Nucl. Fusion* **60**, 046033 (2020).
15. Murari, A. *et al.* Investigating the Physics of Tokamak Global Stability with Interpretable Machine Learning Tools. *Applied Sciences* **10**, 6683 (2020).
16. Vega, J. *et al.* Disruption prediction with artificial intelligence techniques in tokamak plasmas. *Nat. Phys.* **18**, 741–750 (2022).
17. Rattá, G. A., Vega, J. & Murari, A. Improved feature selection based on genetic algorithms for real time disruption prediction on JET. *Fusion Engineering and Design* **87**, 1670–1678 (2012).
18. Yang, Z. *et al.* In-depth research on the interpretable disruption predictor in HL-2A. *Nucl. Fusion* **61**, 126042 (2021).
19. Ferreira, D. R., Martins, T. A. & Rodrigues, P. Explainable deep learning for the analysis of MHD spectrograms in nuclear fusion. *Mach. Learn.: Sci. Technol.* **3**, 015015 (2022).
20. Olofsson, K. E. J., Humphreys, D. A. & Haye, R. J. L. Event hazard function learning and survival analysis for tearing mode onset characterization. *Plasma Phys. Control. Fusion* **60**, 084002 (2018).
21. Pau, A. *et al.* A First Analysis of JET Plasma Profile-Based Indicators for Disruption Prediction and Avoidance. *IEEE Trans. Plasma Sci.* **46**, 2691–2698 (2018).
22. Pau, A. *et al.* A machine learning approach based on generative topographic mapping for disruption prevention and avoidance at JET. *Nucl. Fusion* **59**, 106017 (2019).
23. Aymerich, E. *et al.* A statistical approach for the automatic identification of the start of the chain of events leading to the disruptions at JET. *Nucl. Fusion* **61**, 036013 (2021).
24. Rea, C., Montes, K. J., Pau, A., Granetz, R. S. & Sauter, O. Progress Toward Interpretable Machine Learning–Based Disruption Predictors Across Tokamaks. *Fusion Science and Technology* **76**, 912–924 (2020).
25. Aymerich, E. *et al.* Disruption prediction at JET through deep convolutional neural networks using spatiotemporal information from plasma profiles. *Nucl. Fusion* **62**, 066005 (2022).
26. Chen, C. *et al.* This Looks Like That: Deep Learning for Interpretable Image Recognition. Preprint at http://arxiv.org/abs/1806.10574 (2019).
27. Rudin, C. Stop explaining black box machine learning models for high stakes decisions and use interpretable models instead. *Nat Mach Intell* **1**, 206–215 (2019).

28. Davies, A. *et al.* Advancing mathematics by guiding human intuition with AI. *Nature* **600**, 70–74 (2021).
29. Ribeiro, M. T., Singh, S. & Guestrin, C. 'Why Should I Trust You?': Explaining the Predictions of Any Classifier. Preprint at http://arxiv.org/abs/1602.04938 (2016).
30. Mothilal, R. K., Mahajan, D., Tan, C. & Sharma, A. Towards Unifying Feature Attribution and Counterfactual Explanations: Different Means to the Same End. in *Proceedings of the 2021 AAAI/ACM Conference on AI, Ethics, and Society* 652–663 (2021). doi:10.1145/3461702.3462597.
31. Lundberg, S. & Lee, S.-I. A Unified Approach to Interpreting Model Predictions. Preprint at http://arxiv.org/abs/1705.07874 (2017).
32. Lundberg, S. M. *et al.* Explainable machine-learning predictions for the prevention of hypoxaemia during surgery. *Nat Biomed Eng* **2**, 749–760 (2018).
33. Ke, G. *et al.* LightGBM: A Highly Efficient Gradient Boosting Decision Tree. 9.
34. Zhong, Y. *et al.* Disruption prediction and model analysis using LightGBM on J-TEXT and HL-2A. *Plasma Phys. Control. Fusion* **63**, 075008 (2021).
35. Tay, Y. *et al.* Scaling Laws vs Model Architectures: How does Inductive Bias Influence Scaling? Preprint at http://arxiv.org/abs/2207.10551 (2022).
36. Kates-Harbeck, J., Svyatkovskiy, A. & Tang, W. Predicting disruptive instabilities in controlled fusion plasmas through deep learning. *Nature* **568**, 526–531 (2019).
37. Zhu, J. X. *et al.* Hybrid deep-learning architecture for general disruption prediction across multiple tokamaks. *Nucl. Fusion* **61**, 026007 (2021).
38. Zhu, J. *et al.* Scenario adaptive disruption prediction study for next generation burning-plasma tokamaks. *Nucl. Fusion* **61**, 114005 (2021).
39. Murari, A. *et al.* On the transfer of adaptive predictors between different devices for both mitigation and prevention of disruptions. *Nucl. Fusion* **60**, 056003 (2020).
40. White, R. B., Monticello, D. A. & Rosenbluth, M. N. Simulation of Large Magnetic Islands: A Possible Mechanism for a Major Tokamak Disruption. *Phys. Rev. Lett.* **39**, 1618–1621 (1977).
41. Du, X. D. *et al.* Direct measurements of internal structures of born-locked modes and the key role in triggering tokamak disruptions. *Physics of Plasmas* **26**, 042505 (2019).
42. Carreras, B., Hicks, H. R., Holmes, J. A. & Waddell, B. V. Nonlinear coupling of tearing modes with selfconsistent resistivity evolution in tokamaks. 17 (2014).
43. Pucella, G. *et al.* Onset of tearing modes in plasma termination on JET: the role of temperature hollowing and edge cooling. *Nucl. Fusion* **61**, 046020 (2021).
44. Li, J., Ding, Y., Zhang, X., Xiao, Z. & Zhuang, G. Design of the high-resolution soft X-ray imaging system on the Joint Texas Experimental Tokamak. *Review of Scientific Instruments* **85**, 11E414 (2014).
45. Zhang, X. L., Cheng, Z. F., Hou, S. Y., Zhuang, G. & Luo, J. Upgrade of absolute extreme ultraviolet diagnostic on J-TEXT. *Review of Scientific Instruments* **85**, 11E420 (2014).
46. Greenwald, M. *et al.* A new look at density limits in tokamaks. *Nucl. Fusion* **28**, 2199–2207 (1988).
47. Chen, J., Gao, L., Zhuang, G., Wang, Z. J. & Gentle, K. W. Design of far-infrared three-wave polarimeter-interferometer system for the J-TEXT tokamak. *Review of Scientific Instruments* **81**, 10D502 (2010).
48. Han, D. *et al.* Magnetic diagnostics for magnetohydrodynamic instability research and the detection of locked modes in J-TEXT. *Plasma Sci. Technol.* **23**, 055104 (2021).
49. Sias, G., Cannas, B., Fanni, A., Murari, A. & Pau, A. A locked mode indicator for disruption prediction on JET and ASDEX upgrade. *Fusion Engineering and Design* **138**, 254–266 (2019).
50. Lundberg, S. M. *et al.* From local explanations to global understanding with explainable AI for trees. *Nat Mach Intell* **2**, 56–67 (2020).
51. Friedman, J. H. Greedy function approximation: A gradient boosting machine. *Ann. Statist.* **29**, (2001).
52. Rashmi, K. V. & Gilad-Bachrach, R. DART: Dropouts meet Multiple Additive Regression Trees. Preprint at http://arxiv.org/abs/1505.01866 (2015).
53. *The Shapley value: essays in honor of Lloyd S. Shapley*. (Cambridge University Press, 1988).
54. Slack, D., Hilgard, S., Jia, E., Singh, S. & Lakkaraju, H. Fooling LIME and SHAP: Adversarial Attacks on Post hoc Explanation Methods. in *Proceedings of the AAAI/ACM Conference on AI, Ethics, and Society* 180–186 (ACM, 2020). doi:10.1145/3375627.3375830.

55. Ding, Y. *et al.* Overview of the J-TEXT progress on RMP and disruption physics. *Plasma Sci. Technol.* **20**, 125101 (2018).
56. Wang, N. *et al.* Advances in physics and applications of 3D magnetic perturbations on the J-TEXT tokamak. *Nucl. Fusion* **62**, 042016 (2022).
57. Aledda, R., Cannas, B., Fanni, A., Pau, A. & Sias, G. Improvements in disruption prediction at ASDEX Upgrade. *Fusion Engineering and Design* **96–97**, 698–702 (2015).
58. Raman, R., Lay, W.-S., Jarboe, T. R., Menard, J. E. & Ono, M. Electromagnetic particle injector for fast time response disruption mitigation in tokamaks. *Nucl. Fusion* **59**, 016021 (2019).
59. Grigull, P. *et al.* First island divertor experiments on the W7-AS stellarator. *Plasma Phys. Control. Fusion* **43**, A175–A193 (2001).
60. He, Y. *et al.* Disruption avoidance by stabilizing coupled MHD modes using resonant magnetic perturbations on J-TEXT. (accepted by *Plasma Phys. Control. Fusion*).
61. Shi, P. *et al.* Observation of the high-density front at the high-field-side in the J-TEXT tokamak. *Plasma Phys. Control. Fusion* **63**, 125010 (2021).
62. Shi, P. *et al.* First time observation of local current shrinkage during the MARFE behavior on the J-TEXT tokamak. *Nucl. Fusion* **57**, 116052 (2017).
63. Hu, Q. *et al.* Research on the effect of resonant magnetic perturbations on disruption limit in J-TEXT tokamak. *Plasma Phys. Control. Fusion* **58**, 025001 (2016).